\documentclass[useAMS,usenatbib]{mn2e}
\usepackage{graphicx,lscape}
\usepackage{longtable}

\newcommand{\cmmthree}{\mbox{cm$^{-3}$}}

\newcommand{\kms}{\mbox{km\,s$^{-1}$}}
\newcommand{\degree}{\mbox{$^{\circ}$}}
\newcommand{\msun}{\mbox{M$_{\odot}$}}

\newcommand{\nhthree}{\mbox{NH$_3$}}
\newcommand{\nhone}{\mbox{NH$_3$\,(1,1)}}
\newcommand{\nhtwo}{\mbox{NH$_3$\,(2,2)}}

\newcommand{\water}{\mbox{H$_2$O}}
\newcommand{\vlsr}{\mbox{$v_{\rm LSR}$}}

\def \cm2{\mbox{cm$^{-2}$}}
\def \cm3{\mbox{cm$^{-3}$}}

\title[The HOPS $\nhthree$\,(1,1) and (2,2) catalogues]{The $\water$ southern Galactic Plane Survey
  (HOPS): $\nhone$~and~(2,2) catalogues}

\author[C.\,R.\,Purcell {\it et al.}]{C.\, R.\,Purcell$^{1,2,3}$\thanks{E-mail:\,C.R.Purcell@leeds.ac.uk},
S.\,N.\,Longmore$^{4,5}$, 
A. J. Walsh$^{6}$
M. T. Whiting$^{7}$,
S. L. Breen$^{7}$, \newauthor
T. Britton$^{7,8}$, 
K. J. Brooks$^{7}$, 
M. G. Burton$^{9}$,
M. R. Cunningham$^{9}$, 
J. A. Green$^{7}$, \newauthor
L. Harvey-Smith$^{7}$,
L. Hindson$^{7,10}$,
M. G. Hoare$^2$,
B. Indermuehle$^{7}$,
P. A. Jones$^{9,11}$, \newauthor
N. Lo$^{11,12}$, 
V. Lowe$^{7,9}$, 
C. J. Phillips$^{7}$,
M. A. Thompson$^{10}$, 
J. S. Urquhart$^{7,13}$, \newauthor
M. A. Voronkov$^{7}$ and
G. L. White$^{6}$\\
$^{1}$Sydney Institute for Astronomy (SiFA), School of Physics, The
University of Sydney, NSW 2006, Australia\\
$^{2}$School of Physics \& Astronomy, E.C. Stoner Building, University
of Leeds, Leeds LS2 9JT, UK\\
$^{3}$Jodrell Bank Centre for Astrophysics, School of
Physics and Astronomy, The University of Manchester,
Manchester M13 9PL, UK.\\
$^{4}$European Southern Observatory, Karl Schwarzschildstrasse 2,
Garching 85748, Germany\\ 
$^{5}$Harvard-Smithsonian Centre For Astrophysics, 60 Garden Street,
     Cambridge, MA, 02138, USA\\
$^{6}$Centre for Astronomy, School of Engineering and Physical
     Sciences, James Cook University, Townsville, QLD 4811, Australia\\
$^{7}$CSIRO Astronomy and Space Science, PO BOX 76, Epping, NSW 1710,
     Australia\\
$^{8}$Department of Physics \& Astronomy, Faculty of Science,
     Macquarie University, NSW 2109, Australia\\
$^{9}$School of Physics, University of New South Wales, Sydney, NSW
     2052, Australia\\
$^{10}$Centre for Astrophysics Research, Science and Technology
     Research Institute, University of Hertfordshire, AL10 9AB, UK\\
$^{11}$Departamento de Astronom\'ia, Universidad de Chile, Casilla 36-D,
     Santiago, Chile\\
$^{12}$Laboratoire AIM Paris-Saclay, CEA/Irfu - Uni. Paris Did\'erot -
     CNRS/INSU, 91191 Gif-sur-Yvette, France\\
$^{13}$Max-Planck-Institut f\"{u}r Radioastronomie, Aug dem H\"{u}gel
     69, 53121 Bonn, Germany}
\begin{document}

\date{2012-Jul-25}

\pagerange{\pageref{firstpage}--\pageref{lastpage}} \pubyear{2011}

\maketitle

\label{firstpage}

\begin{abstract}
The {\bf H}$_2${\bf O} Southern Galactic {\bf P}lane {\bf
S}urvey (HOPS) has mapped a 100\,degree strip of the Galactic
plane ($-70\degree>l>30\degree$, $|b|<0.5^{\circ}$) using the 22-m
Mopra antenna at 12-mm wavelengths. Observations were conducted 
in on-the-fly mode using the Mopra spectrometer (MOPS), targeting
water masers, thermal molecular emission and radio-recombination
lines. Foremost among the thermal lines are the 23\,GHz transitions of
\nhthree~J,K\,=\,(1,1) and (2,2), which trace the densest parts of 
molecular clouds ($n>10^4$\,cm$^{-3}$). In  this paper we present the
\nhthree\,(1,1) and (2,2) data, which have a resolution of 2~arcmin and
cover a velocity range of $\pm200\,\kms$. The median sensitivity of
the $\nhthree$ data-cubes is $\sigma_{T_{\rm mb}}=0.20\pm0.06$\,K. For
the (1,1) transition this sensitivity equates to a 3.2\,kpc distance
limit for detecting a 20\,K, 400\,$\msun$ cloud at the 5$\sigma$
level. Similar clouds of mass 5,000\,$\msun$ would be detected as far
as the Galactic centre, while 30,000\,$\msun$ clouds would be seen
across the Galaxy. We
have developed an automatic emission finding procedure based on the ATNF
{\scriptsize DUCHAMP} software and have used it to create a new 
catalogue of 669 dense molecular clouds. The catalogue is 100 percent
complete at the 5$\sigma$ detection limit ($T_{\rm mb}=1.0$\,K). A
preliminary analysis of the ensemble cloud properties suggest that the
near kinematic distances are favoured. The cloud positions 
are consistent with current models of the Galaxy containing a long
bar. Combined with other Galactic plane surveys this new
molecular-line dataset constitutes a key tool for examining Galactic
structure and evolution. Data-cubes, spectra and catalogues are
available to the community via the HOPS website.
\end{abstract}



\section{Introduction}
HOPS ({\bf H}$_2${\bf O} Southern Galactic {\bf P}lane {\bf
  S}urvey) is a project utilising the Mopra radio
telescope\footnote{Mopra is a 22-m single dish mm-wave telescope
  situated near Siding Spring mountain in New South 
  Wales, Australia.} to simultaneously map spectral-line
emission along the southern Galactic plane across the full 12-mm
band (frequencies of 19.5 to 27.5\,GHz). Since the survey began in
2007 \citep{walsh2008} HOPS has mapped 100 square degrees of the Galactic
plane, from $l=290^\circ$, continuing through the Galactic centre 
to $l=30^\circ$ and with Galactic latitude $|b|\leq
0.5^\circ$. The aim of the survey is to provide an untargeted census
of 22.235\,GHz $\water$ (6$_{16}-$5$_{23}$) masers and thermal line
emission towards the inner Galaxy. Observations were completed in
2010 and the survey properties, observing parameters, data reduction
and $\water$ maser catalogue are described in \citet{walsh2011}
(hereafter Paper~I). 

The primary thermal lines targeted by HOPS are those of ammonia
($\nhthree$), whose utility as a molecular thermometer in a broad
range of environments is unsurpassed. With an effective critical
density of $\sim10^4$\,$\cm3$ \citep{ho_townes1983}, $\nhthree$ traces
dense molecular gas and it is often associated with the hot molecular
core phase of high-mass star formation (e.g., \citealt{L07A},
\citealt{morgan2010}), where it exhibits temperatures in excess of
30\,K. $\nhthree$ is excited in gas with kinetic temperatures greater
than $\sim5$\,K \citep{pickett1998} and is also found associated with
cool ($T<10$\,K) dense clouds. Such regions are too cold for more
common gas tracers, like CO, to remain in the gas phase. Instead they
are frozen out onto 
the surfaces of dust grains \citep{bergin2006}. The
J,K\,=\,(1,1) inversion transition exhibits prominent hyperfine
structure, which can be used to infer the optical depth of 
the transition. In clouds forming high-mass stars ($\ge8\,\msun$)
and under optically thin conditions, the peak brightness of the four
groups of satellite lines is approximately half that of the 
central group \citep{rydbeck1977}. Comparison of the (1,1) and
higher J,K inversion transitions can be used to estimate the
rotational temperature of the gas.

\begin{figure*}
  \centering
  \includegraphics[ angle=90, height=22.6cm, trim=0 0 0 0]{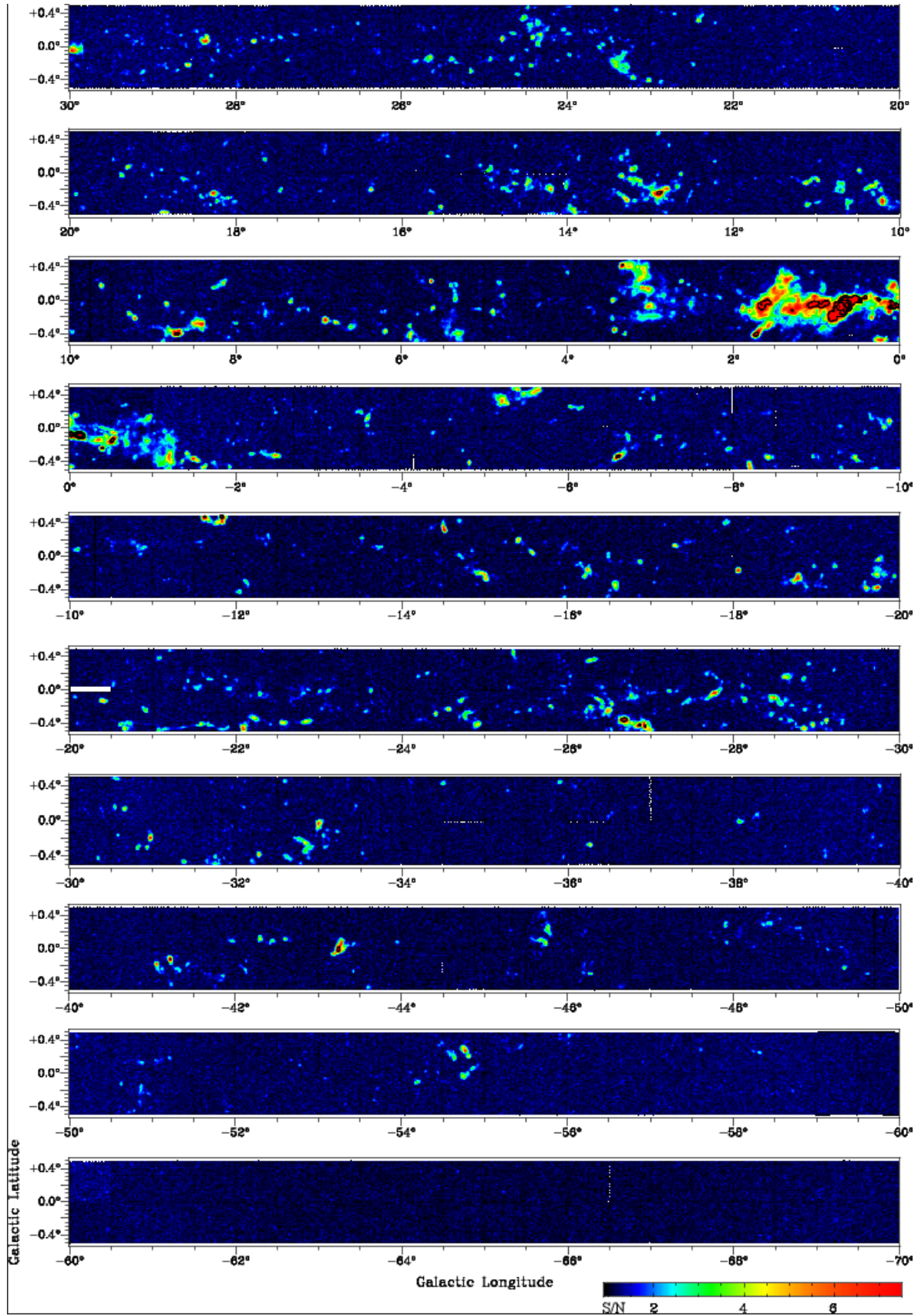}
  \caption{Map of the peak \nhthree\,(1,1) signal-to-noise (S/N)
    across the HOPS survey area. The map was made by smoothing the S/N
    cube spatially (final beam FWHM\,=\,2.5~arcmin) and in velocity
    (hanning-window\,=\,5) and measuring the peak brightness in each
    spectrum. An approximate brightness temperature scale may be found
    by multiplying by the average root-mean-squared noise temperature
    $\langle\sigma_{T_{\rm mb}}\rangle$\,=\,0.20\,K.}
  \label{fig:s2n_full}
\end{figure*}
\begin{figure*}
  \centering
  \includegraphics[ angle=90, height=22.6cm, trim=0 0 0 0]{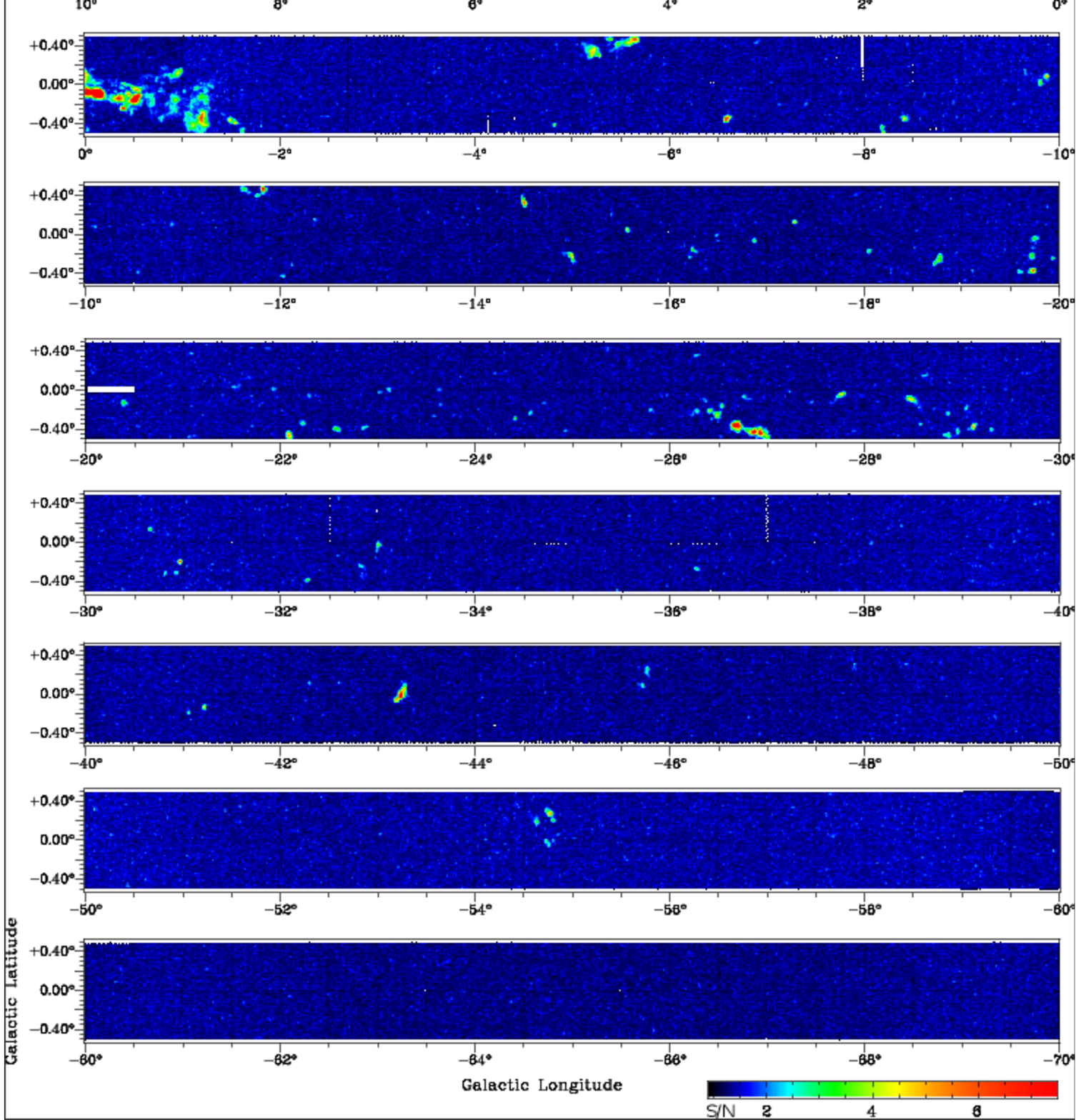}
  \caption{Map of the peak \nhthree\,(2,2) signal-to-noise (S/N)
    across the HOPS survey area. }
  \label{fig:s2n_full_22}
\end{figure*}

In this second paper we present the HOPS $\nhone$ and $\nhtwo$
datasets and the automatic finding procedure used to create
catalogues of emission. We illustrate the basic properties of the
catalogues, which form the basis for further analysis. A third paper
(Longmore et al. {\it in prep.}) will published the properties of the
catalogue derived by fitting the $\nhone$ and $\nhtwo$ spectra with
model line profiles (i.e. temperature, density, mass and evolutionary
state).


\section{Observations and data reduction}
The HOPS observations and survey design are described in
\cite{walsh2008} and in Paper~I. For convenience a brief summary is
provided here. 


\subsection{Observations}
Observations were conducted using the 22-m Mopra radio-telescope
situated at latitude 31:16:04 south, longitude 149:05:58 east and at an
elevation of 850 meters above sea level. Data were recorded over four
summer seasons during the years 2007\,--\,2010 using a
single-element 12-mm receiver.

The digitised signal from the receiver was fed into the Mopra Spectrometer
(MOPS) backend, which has an 8.3\,GHz total bandwidth split into four
overlapping sub-bands, each 2.2\,GHz wide. During
the HOPS observations the spectrometer was configured in `zoom' mode,
whereby each of the sub-bands contained four 137.5\,MHz zoom
bands. Up to sixteen zoom bands may be recorded to disk, each split into
4096 channels. A single zoom band covered both the J,K\,=\,(1,1) 
and (2,2) $\nhthree$ inversion transitions, which have line-centre rest
frequencies of 23.6944803\,GHz and 23.7226336\,GHz,
respectively \citep{pickett1998}. This frequency setup resulted in a
velocity resolution of $\sim0.42\,\kms$ per channel, spanning a
velocity range of $1739\,\kms$.

At 23.7\,GHz the full-with half-maximum (FWHM) of the Mopra
beam is 2.0~arcmin \citep{urquhart2010}, meaning that the 100 square
degree target area would require 90,000 pointings to create a fully
sampled point-map. Based on experience gained during the pilot observations
\citep{walsh2008} the survey area was divided up into 400
$0.5\degree\times0.5\degree$ maps in Galactic coordinates. The
telescope was driven in on-the-fly (OTF) mode, raster-scanning in
either Galactic $l$ or $b$ and recording spectra every six seconds. At the end
of each row spectra were taken of an emission-free reference position
situated away from the Galactic plane. Scan rows were offset by half a
beam FWHM and a scan rate was chosen to ensure Nyquist sampling at the
highest frequency. Each map was observed twice by scanning in orthogonal
$l$ and $b$ directions the results were averaged together to minimise
observing artifacts. 


\begin{figure}
  \centering
  \includegraphics[width=8.0cm, angle=0, trim=0 0 0 0]{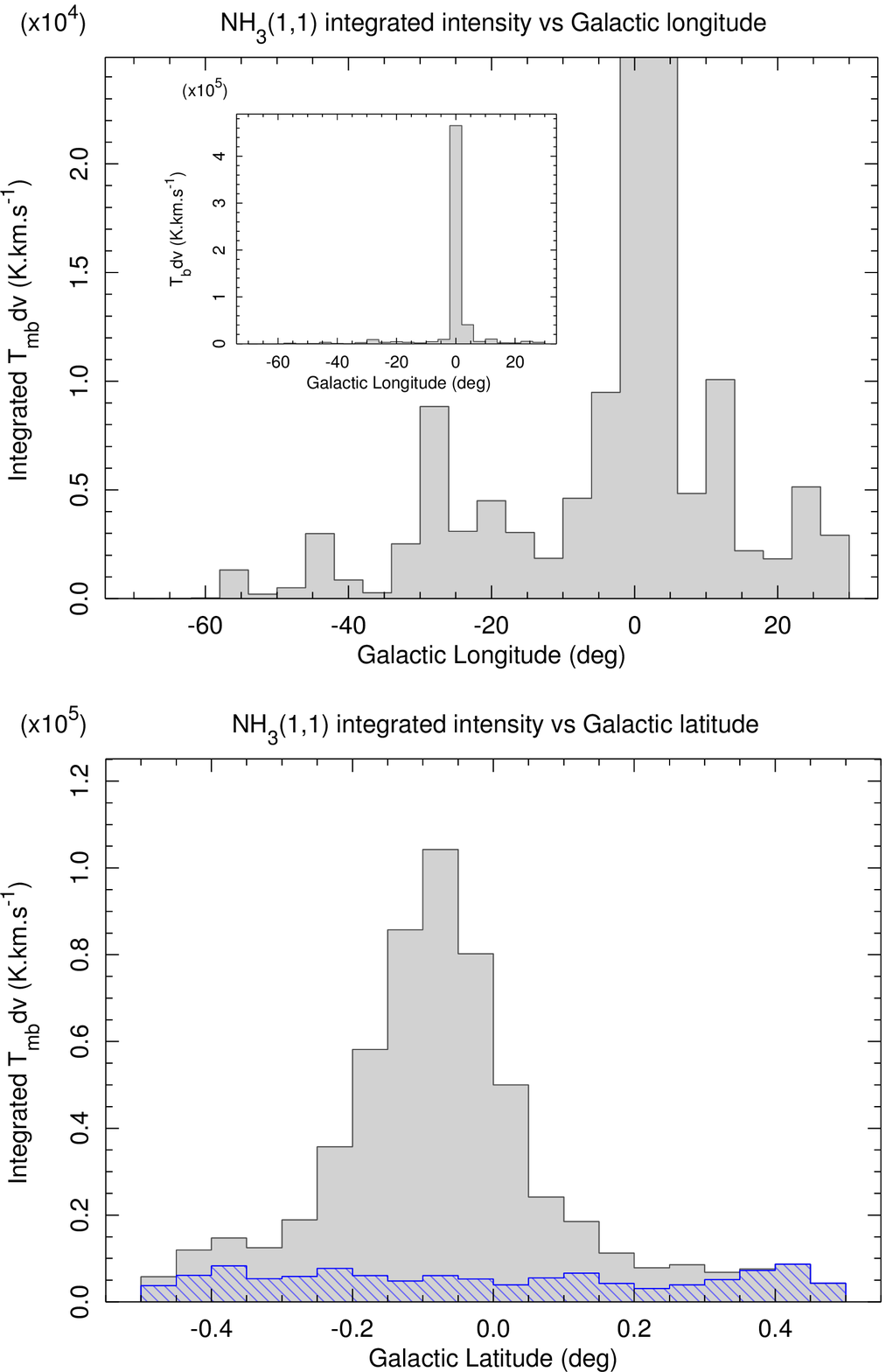}
  \caption{The top panel shows the sum of the \nhthree\,(1,1)
    integrated intensity over the survey region as a function of
    Galactic longitude in 
  four-degree bins. The longitude bin covering the Central Molecular
  Zone (CMZ, $-2^{\circ}<\,l<2^{\circ}$) peaks at
 $\int T_{\rm mb}\,$dv\,=\,$4.65\times10^5$\,K\,$\kms^{-1}$ and the y-axis
 scale has been truncated so as to show the remainder of the
 emission. The inset shows how the plot appears scaled to the full
 $\int T_{\rm mb}\,$dv range. In the bottom panel is a similar plot of
 $\int T_{\rm mb}\,$dv versus Galactic latitude. The filled grey histogram
 includes all emission, while the hatched histogram includes only
 emission outside the CMZ.}
  \label{fig:Tbdv_vs_lb}
\end{figure}

\begin{figure}
  \centering
  \includegraphics[height=8.cm, angle=-90, trim=0 0 0 0]{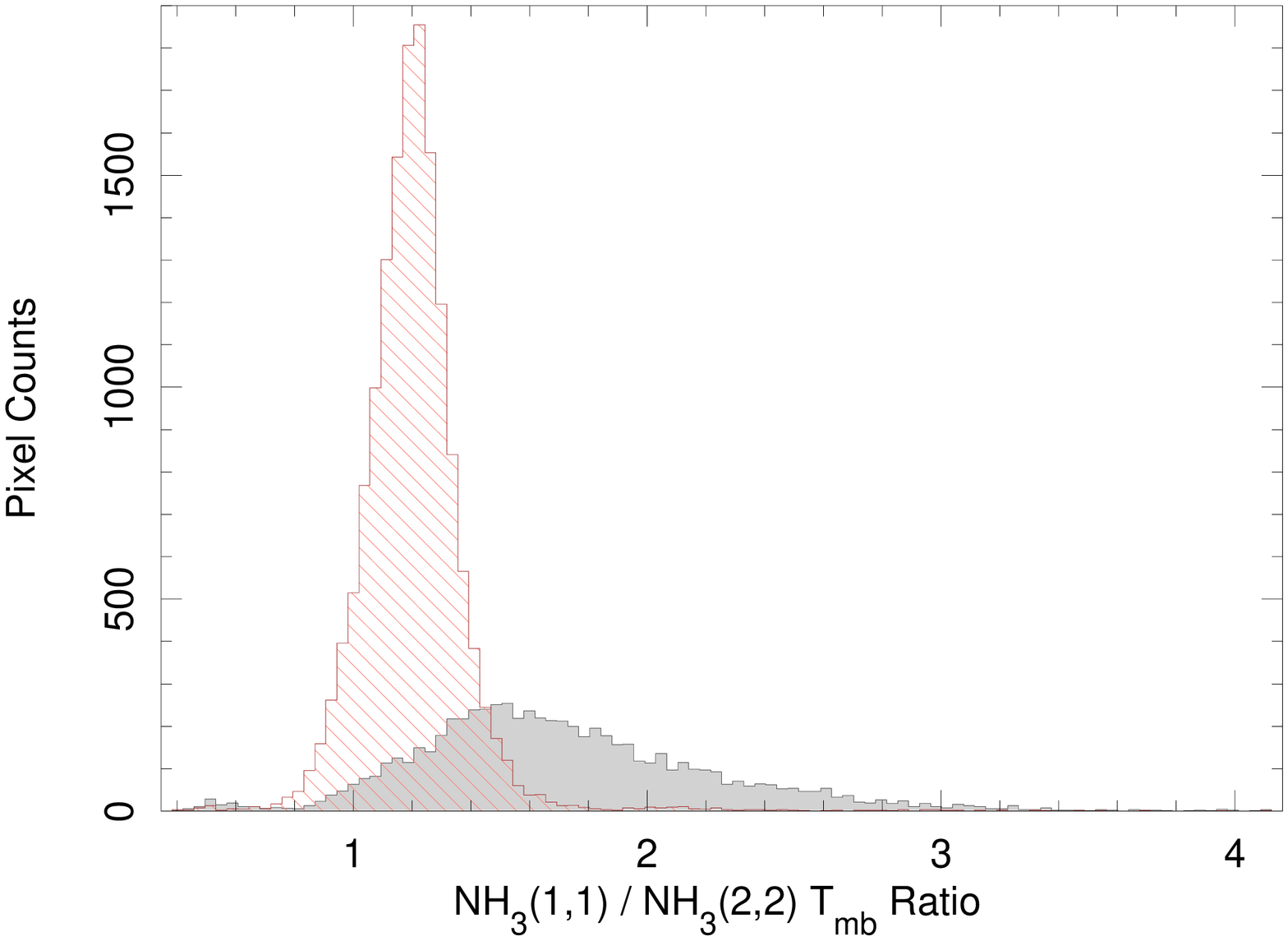}
  \caption{Plots of the brightness ratio $T_{\rm mb,\,(1,1)}/T_{\rm
      mb,\,(2,2)}$ for every emitting voxel ($l$-$b$-$v$ data element) with
    $T_{\rm mb}>3\sigma$ in both HOPS $\nhthree$ datasets. The
    hatched histogram includes only voxels from the CMZ
    ($|l|<2\degree$) while the solid histogram includes the remaining
    emission from the outer Galaxy.} 
  \label{fig:Tmb_ratio}
\end{figure}

\subsection{Data reduction}\label{sec:reduction}
Raw data from the telescope were written into {\scriptsize RPFITS}
format files consisting of lists of spectra with associated
time-stamps and coordinate information. The ATNF {\scriptsize
  LIVEDATA}\footnote{http://www.atnf.csiro.au/people/mcalabre/livedata.html} 
software package was used to apply a bandpass correction by forming a quotient
between the reference spectrum and individual spectra in its
associated scan row. The software was configured to fit and subtract a
first order polynomial from the bandpass, excluding any strong lines
and 150 noisy channels at each edge.

The {\scriptsize GRIDZILLA} package was used to resample the
bandpass corrected spectra onto a regular coordinate grid. The
software also performed interpolation in velocity space to convert
measured topocentric frequency channels into LSR velocity. At this
stage spectra with system temperatures $T_{\rm sys}>120$\,K were
discarded. Software settings were chosen to produce 
data cubes with pixel scales of $30\times30$ arcsec and a resolution of
2~arcmin. The final data products were $1\times10$~deg$^2$ cubes with
3896 usable velocity channels. For each spectral line within this 
pass-band the data has been cropped to $\pm\,200\,\kms$
centred on the line centre velocity (determined from the line rest
frequency), which corresponds to the normal velocity range for
molecular material in the Galaxy \citep{dame2001}.

A small number of spectra ($\sim10$ percent) were affected by a
baseline ripple or a depressed zero level. We fitted additional
polynomial baselines of order 3\,--\,5 to the line-free channels of
these data. See Appendix~\ref{sec:baselines} for more detail.


\subsection{Temperature scale and uncertainty}
Data from Mopra are calibrated in brightness temperature units (K) on
the $T_{\rm A}^*$ scale, i.e., they are corrected for radiative loss and
rearward scattering (although not atmospheric attenuation at 12-mm
wavelengths - see \citealt{kutnerulich1981} for a review and
also \citealt{ladd2005}). It is desirable to convert this onto a
telescope-independent main beam brightness temperature scale ($T_{\rm
  mb}$), which is also corrected for forward scattering. The measured
$T_{\rm mb}$ of a source would equal the true brightness temperature
if the source just filled the main beam. \citet{urquhart2010}
characterised the telescope efficiencies between 17\,GHz and 49\,GHz
and we divided the HOPS data by the main-beam efficiency, $\eta_{\rm
  mb}$, given therein. For both \nhthree\,(1,1) and (2,2) data
$\eta_{\rm mb}=0.571$. The uncertainty on the $T_{\rm mb}$ scale for
HOPS is $\sim30$ percent \citep{walsh2011}. 


\section{The \nhthree~data}
In contrast to other large molecular line surveys of the Galactic
plane, which use relatively abundant CO isotopologues
(e.g. \citealt{dame2001}, \citealt{jackson2006}) HOPS has focused on
the inversion-rotation transitions of \nhthree.  Detections derive
from the densest parts of giant molecular clouds where gas is
condensed into cool clumps, or in the case of warm gas ($T>30$\,K)
surrounding centrally heated star-forming regions (e.g.,
\citealt{Purcell2009}).


\subsection{$\nhthree$ emission properties}
Figures~\ref{fig:s2n_full} and~\ref{fig:s2n_full_22} present peak
signal-to-noise (S/N) maps covering the \nhthree\,(1,1) and (2,2)
passbands, respectively ($|v_{\rm LSR}|<200\,\kms$). Such maps are the
easiest way to visualise the distribution of emission on the sky
without becoming confused by noise-related artifacts. The maps were
constructed by smoothing a S/N cube to a spatial resolution of 
2.5~arcmin and then hanning-smoothing in the spectral dimension using a
filter width of five channels ($\sim2.0\,\kms$). For any given spatial
pixel the brightest spectral channel was recorded on the map. The S/N
cube was generated during the emission-finding process, which is
explained in Section~\ref{sec:source_detect_extract}. 


\begin{figure}
  \centering
  \includegraphics[width=8.2cm, angle=0, trim=0 0 0 0]{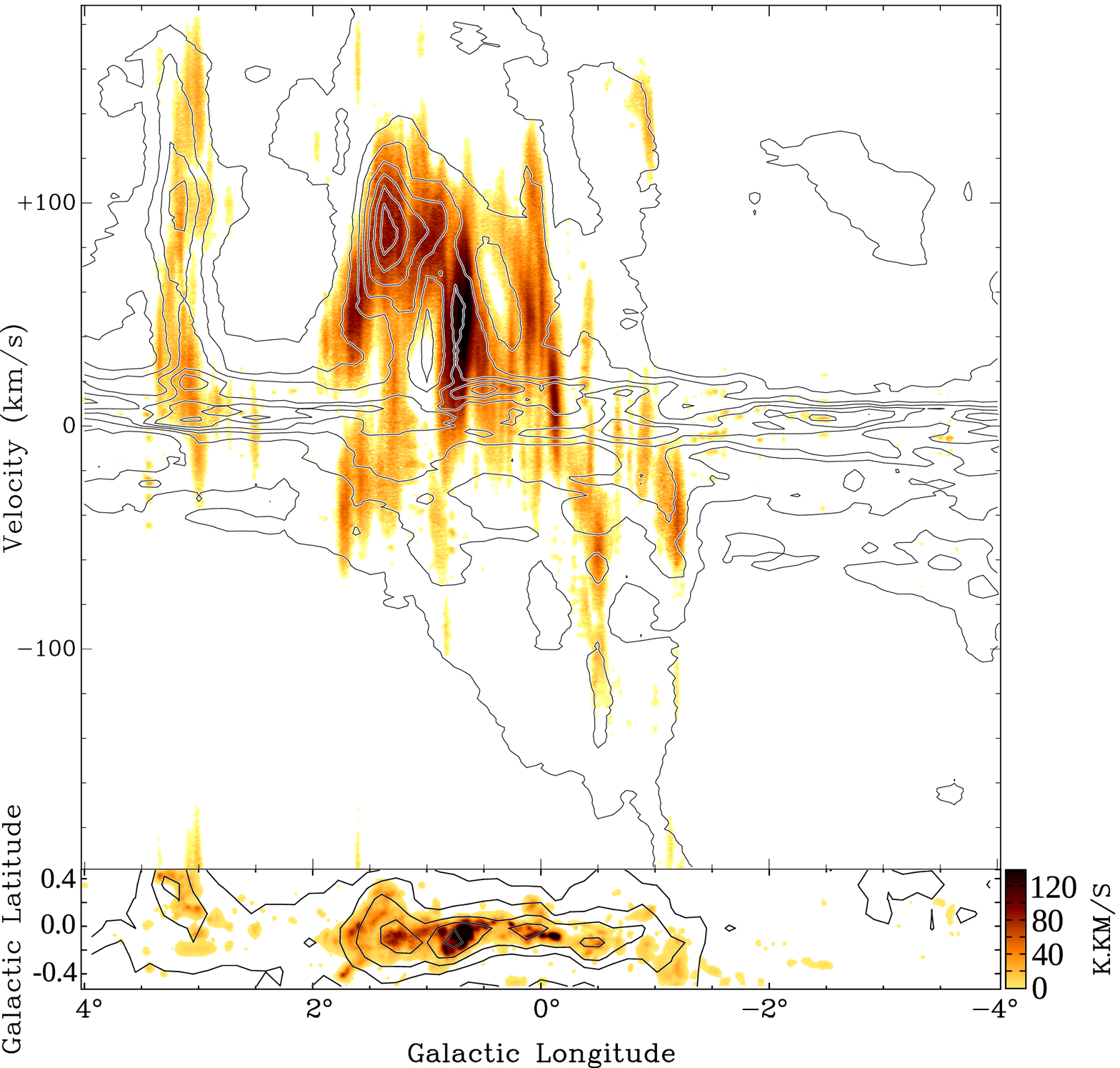}
  \caption{Detail of the longitude-velocity diagram showing the
    Galactic centre region (top). \nhthree\,(1,1) emission is shown in
    colourscale and the \citet{dame2001} CO\,(1\,--\,0) contours are
    overplotted. While the overall emission morphology of both
    molecules is similar, the brightest peaks are offset in $l$ and
    $v$. The emission seen at $v<-170\,\kms$ is due to broad linewidth
    \nhthree\,(2,2) emission encroaching on the \nhthree\,(1,1)
    pass-band. The bottom panel shows the \nhthree\,(1,1) integrated
    intensity map for the same $l$ range.}  \label{fig:PV_GC} 
\end{figure}

\begin{figure}
  \centering
  \includegraphics[width=8.7cm, angle=0, trim=0 0 0 0]{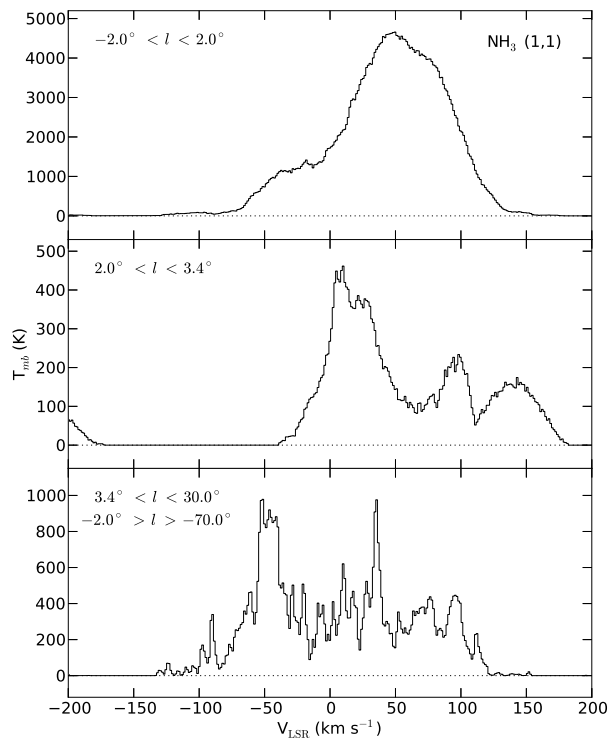}
  \caption{Integrated \nhthree\,(1,1) spectrum for the Galactic centre
    (top panel), the region around the G$3.34+0.42$ cloud, also known
    as Bania's Clump 2 (middle panel), and the outer Galaxy (bottom
    panel). Spectra were integrated over the full latitude range
    ($\pm0.5\degree$).} 
  \label{fig:integ_spec_NH3_11}
\end{figure}

\subsubsection{Galactic distribution of emission}
The \nhthree\,(1,1) map (Figure~\ref{fig:s2n_full}) reveals a
multitude of distinct clouds\footnote{We refer colloquially to any
  contiguous region of emission as a `cloud' without implying
  membership of any object category based on size-scale.} between
$-30^{\circ}\,<l\,<\,30^{\circ}$. The density of clouds falls off
rapidly in the southern Galaxy beyond $l<-30^{\circ}$. Several well 
known star-formation complexes are detected as bright peaks
including G333.0-0.4 (\citealt{bains2006}, \citealt{Wong2008},
\citealt{lo2009}), W43 (G29.9, \citealt{Dame1986},
\citealt{NguyenLuong2011}), G305 (\citealt{Clark2004},
\citealt{Hindson2010}) and the `Nessie' filament at
$l=-21.5^{\circ}\,$, $b=-0.48\degree$  
\citep{jackson2010}. Emission from the Galactic centre region is
most prominent between $-2^{\circ}\,<l\,<\,2^{\circ}$ and although
symmetric in overall morphology, is significantly brighter at positive
longitudes. Gas in this region, known in the literature as the Central
Molecular Zone (CMZ), is characterised by high temperatures and
densities, and by large velocity dispersions as it flows into the
central 200\,pc of the Galaxy \citep{Morris1996}.

Figure~\ref{fig:Tbdv_vs_lb} illustrates the distribution 
of \nhthree\,(1,1) emission as histogram. The top panel shows
the total integrated intensity (summed over all $v$ and $b$) as  
a function of Galactic longitude in 4$\degree$ bins. The CMZ contains
80.6 percent of the detected \nhthree\,(1,1) emission and dominates
the plot, which is scaled to 
show the less intense emission in the outer Galaxy. The bottom
panel shows the same plot as a function of Galactic latitude in 3~arcmin
bins. The filled grey histogram 
incorporates all emission in the HOPS target region. The broad and
intense peak at $b=-0.075\degree$ is solely due to emission from the CMZ.
However, with the Galactic centre removed (hatched histogram),
the distribution with $b$ is relatively flat. 

It is interesting to note that the Galactic distribution of H$_2$O
masers found in HOPS \citep{walsh2011} is very different to that
of the dense molecular gas traced by $\nhthree$. The masers show no
evidence of a longitude peak at the Galactic centre, but are clustered
around the mid-plane of the Galactic disk with an angular scale-height of
$\theta_{\rm FWHM}=0.60^{\circ}$. The implications for star-forming
environments in the CMZ compared to the spiral arms will be discussed
in a future paper (Longmore et al., {\it in prep}).


\subsubsection{\nhthree\,(1,1) versus \nhthree\,(2,2) emission}
Only one-third of the \nhthree\,(1,1) emission regions detected above
3$\sigma$ have counterparts in the \nhthree\,(2,2) data, despite the
data having similar sensitivities. Most clouds with $T_{\rm 
  mb,\,(1,1)}>$1\,K are detected in the (2,2) data, although there
are exceptions (for example the cloud at $l=-2.5\degree,
b=0.33\degree$). Where emission is detected in both lines, the
brightness ratio $T_{\rm mb,\,(1,1)}/T_{\rm mb,\,(2,2)}$ varies
between $\sim0.2$ and $\sim4.2$ over the HOPS area. Clouds with low
ratios or no detected \nhthree\,(2,2) emission tend to be associated with
extinction features in the GLIMPSE\footnote{Galactic Legacy Infrared
  Mid-Plane Survey Extraordinaire http://www.astro.wisc.edu/sirtf/}
$8\,\micron$ infrared images and are excellent candidates for the cold
and dense precursors to clusters of high-mass stars. Strikingly, the 
morphology of the CMZ is almost identical in both
datasets and the brightness-temperature ratio is uniformly close to
one. Figure~\ref{fig:Tmb_ratio} illustrates this difference
graphically by plotting the distribution of brightness-temperature
ratios between every common emitting voxel (3-dimensional data
elements in $l$, $b$ \& $v$) with $T_{\rm mb}>3\sigma$. The tall,
narrow histogram (hatched) contains voxels drawn from the CMZ only and
has a median of 1.2 and a width of $\sigma=0.19$. By comparison the data 
from the outer Galaxy has a median of 1.6 and $\sigma=0.50$. Lower
ratios in the CMZ are consistent with warmer excitation conditions,
but may also be due in part to the effects of high optical depths or
line saturation.

A small proportion of \nhthree\,(2,2) emission is not associated with
\nhthree\,(1,1) emission at the same location. The J,K\,=\,(1,1)
transition is more easily excited than the (2,2), hence such
\nhthree\,(2,2) detections are most likely artifacts. We analyse the
completeness and spurious source counts in
Sections~\ref{sec:nh3_catalogue} and~\ref{sec:completeness}.


\begin{figure*}
  \centering
  \includegraphics[angle=90, height=22.6cm, trim=0 0 0 0]{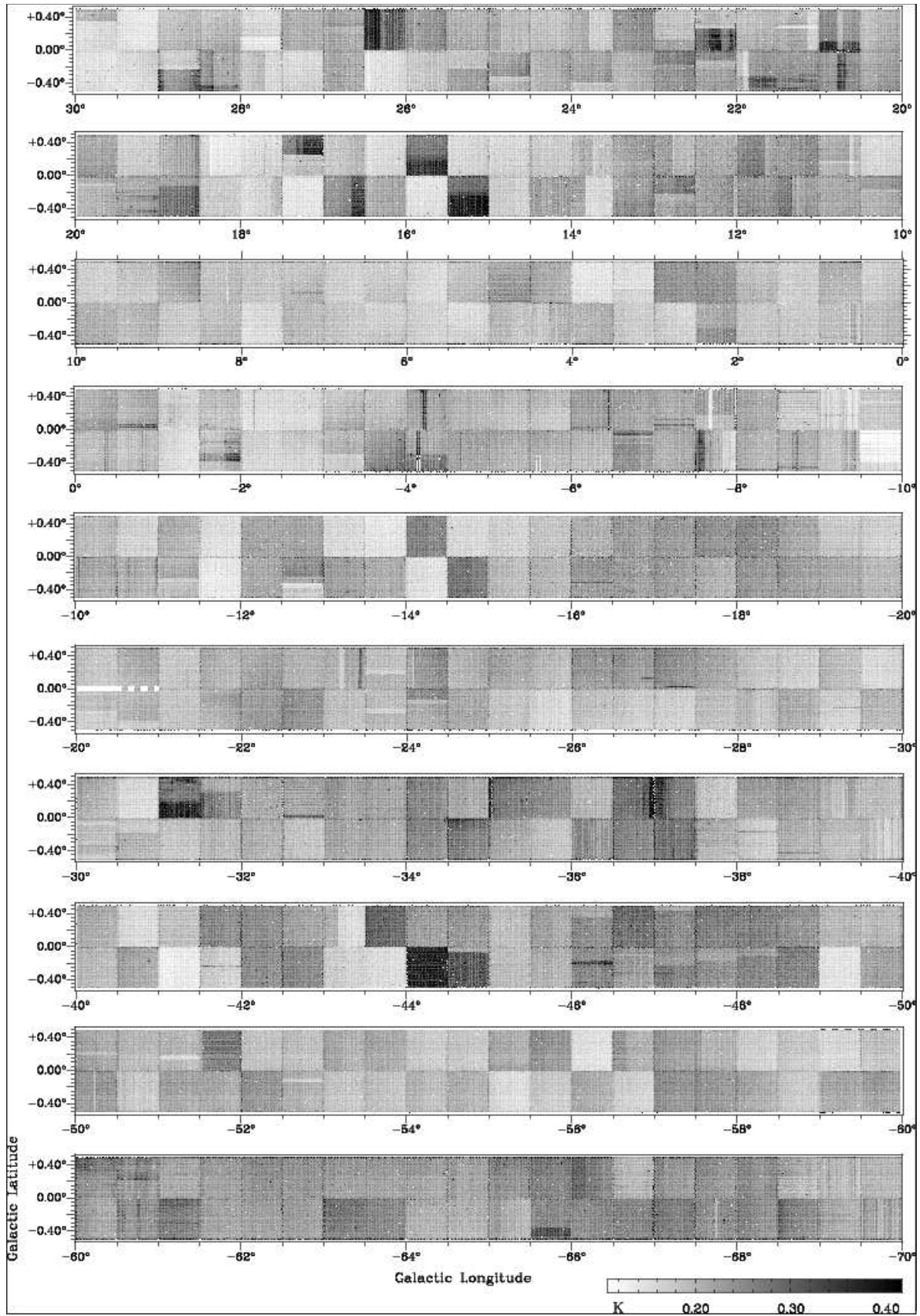}
  \caption{Image of the root-mean-squared noise temperature in $T_{\rm
    mb}$ units for the entire HOPS region made from the
    \nhthree\,(1,1) data. Individual observing blocks are visible as
    0.5$\degree$ chequered squares, while scanning artifacts present
    as lines of high or low noise in Galactic $l$ and $b$. The median
    root-mean-squared noise is 0.20\,K\,$\pm$\,0.05\,K, where the
    uncertainty is the FWHM of the total noise distribution (also see
    Figure~\ref{fig:noise_hist}.}
  \label{fig:noise_full}
\end{figure*}

\subsubsection{Position-velocity Diagram}
Spectral lines originating from the Galactic centre region have
extremely broad profiles ($>50\,\kms$) and merge together into a
single bright region of emission with a complex morphology.
Figure~\ref{fig:PV_GC} (top) is an $l$-$v$ diagram showing the inner
eight degrees of the Galactic plane and the prominent CMZ. To make the
map, voxels without detected emission were set to zero and the data
were summed in Galactic latitude. The
corresponding integrated intensity map is presented in the bottom
panel. A significant velocity gradient exists across the 
emission, which has also been observed in the \citet{dame2001}
CO\,(1\,--\,0) data. \citet{RodriguezFernandez2008} have modelled the
CMZ as the gas response to a short `nuclear bar', which itself is
embedded within a longer Galactic bar, thought to be inclined towards
us at an angle of $25\degree\,-\,35\degree$, with the nearer end at
positive Galactic longitudes. The large broad-line region of emission
at $l=3.1\degree$ is known as Bania's Clump 2 \citep{Bania1977} and
has been theorised 
to lie at the closest end of innermost elongated x-1 orbit around the
Galactic centre \citep{Bally2010}. The CO\,(1\,--\,0) contours from
the 12~arcmin resolution \citet{dame2001} survey are overlaid on the
$l$-$v$ diagram for comparison. The overall morphology of the emission
is similar, 
however, the 2~arcmin resolution $\nhthree$ data shows more 
structure, both spatially and spectrally. Galactic CO\,(1\,--\,0)
emission is optically thick compared to $\nhthree$, which probes the
kinematics of the molecular clouds at greater depths. $\nhthree$
detections in HOPS are severely distance limited (see
Section~\ref{sec:completeness_illustrated}) and detecting emission
from beyond the Galactic centre is difficult, even in the highly
excited central Galaxy.


\subsubsection{Integrated spectrum}
The \nhthree\,(1,1) spectrum is characterised by five groups of
emission lines seperated $\sim10\,\kms$. Under optically thin
conditions the central group is approximately twice the brightness of
the satellites. Figure~\ref{fig:integ_spec_NH3_11} presents
\nhthree\,(1,1) spectra summed over all $b$ and selected $l$ ranges.
In the top panel is shown the integrated spectrum of the CMZ between
$-2\degree<l<2\degree$. Individual spectra (and line-groups within
spectra) are blended into a single emission feature spanning
$-160\,\kms<v_{\rm LSR}<160\,\kms$. The middle panel shows the
integrated spectrum of the region dominated by Bania's clump 2
(G$3.34+0.42$ in this work), which has the broadest linewidths outside
of the CMZ. Emission at $v_{\rm LSR}<-170\,\kms$ is due to broad
\nhthree\,(2,2) lines impinging on the \nhthree\,(1,1) 
bandpass. The spectrum in the bottom panel is constructed from the 
outer Galaxy data only and contains multiple distinct velocity
features deriving from individual regions of emission. The peaks at
approximately $-50\,\kms$ and $+40\,\kms$ are likely associated with
the Galactic ring and Scutum-Centarus spiral arm, respectively (see
Section~\ref{sec:gal_struct}). Some of the complexity evident in the
ensemble spectrum may also be attributed to the summation of satellite
line-groups. 


\subsection{Noise characteristics}
Figure~\ref{fig:noise_full} shows an image of the root-mean-squared (RMS)
noise temperature $\sigma_{T_{\rm mb}}$ over the whole survey region. The
image was constructed by measuring the RMS noise along the spectral
dimension of the $\nhone$ cube after masking off all channels containing
line-emission. We are confident that the noise map does not contain any
contribution from real emission as contamination from the
broad-linewidth CMZ is absent, and this represents the worst-case
scenario. 

The defining feature of Figure~\ref{fig:noise_full} is the chequered
pattern corresponding to individual $0.5\times0.5$~degree
maps. The variable noise level between maps is due to temporal changes
in the conditions encountered while observing. Within each map,
rapidly fluctuating cloud cover and flagged or missing spectra lead to
high-noise `hot-spots'. More gradual changes in observing conditions
(e.g., humidity and elevation) manifest themselves as significant
noise variations between scan-rows in $l$ or $b$. Automatically
finding line-emission in such an inhomogeneous data-cube poses a
distinct challenge.

\begin{figure}
  \centering
  \includegraphics[width=8.3cm, angle=0, trim=0 0 0 0]{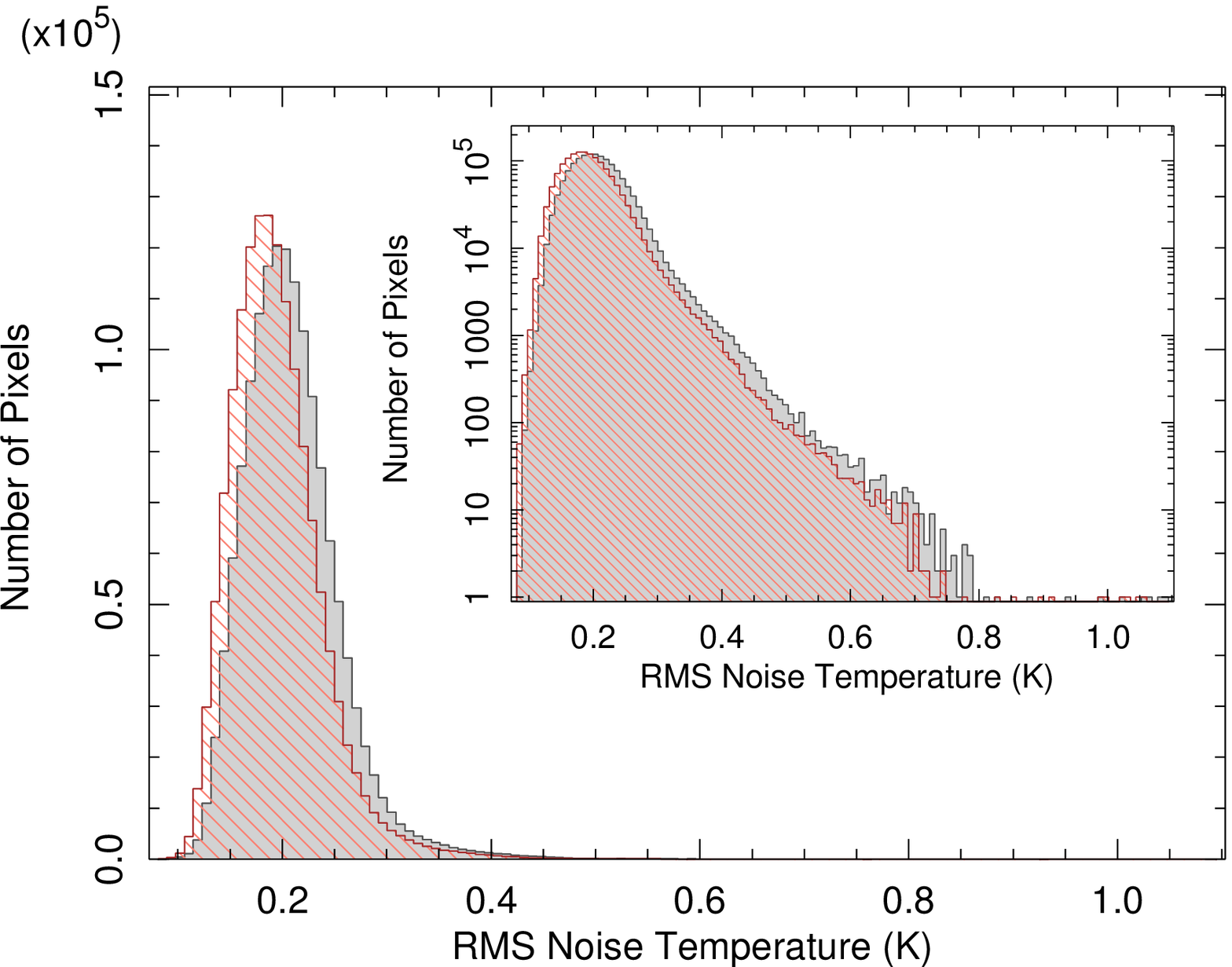}
  \caption{Histograms of the RMS noise temperature for all $(l,b)$
    spatial pixels in the \nhthree\,(1,1) and (2,2) data (grey-shaded and
    hatched histograms, respectively). The noise temperatures
    were determined from the line-free channels in the spectra. Inset is the
    same histogram, but with a logarithmic scaling on the Y-axis to
    enhance the view of the high-noise tail.}
  \label{fig:noise_hist}
\end{figure}

Histogram plots of the noise temperature $\sigma_{T_{\rm mb}}$ for all
spatial pixels in the HOPS \nhthree\,(1,1) and (2,2) data are plotted
in Figure~\ref{fig:noise_hist}. Both datasets have Gaussian shaped
distributions, except for a small number of pixels which occupy a
tail extending to higher noise temperatures (0.71\,\% above
3$\sigma$).  The median noise temperature of the J,K\,=\,(1,1)
data is higher than the (2,2): $\sigma_{T_{\rm mb\,\,(1,1)}}=0.201$\,K
compared to $\sigma_{T_{\rm mb,\,(2,2)}}=0.186$\,K, although the
Gaussian part of both distributions have similar widths at
FWHM$_{(1,1)}=0.056$\,K and FWHM$_{(2,2)}=0.053$\,K. Upon inspection of
the data we find that the high-noise pixels in the tail arise from
$\sim10\degree$ maps observed during poor weather and from the edges of the
maps where fewer raw spectra contribute to individual pixels. The
difference between the \nhthree\,(1,1) and (2,2) distributions stems
from a single low-noise \nhthree\,(2,2) map which had additional data
added into the processing pipeline. No emission was detected in this
map, which lies in the far southern Galactic plane at $l=-60.25$, $b=0.25$.

\section{Source finding and measurement}\label{sec:source_detect_extract}
The full HOPS dataset covers 100 square degrees of the sky, equivalent
to over 90,000 overlapping Gaussian beams at 23.7\,GHz. For each position
we obtain a spectrum spanning a 400\,$\kms$ velocity range, or 931
channels. Manually identifying emission in such a large volume of data
would be prohibitively tedious and likely error-prone, so an automatic
emission finding procedure was written for HOPS. At the core of the
method is the ATNF's {\sc 
  duchamp}\footnote{http://www.atnf.csiro.au/people/Matthew.Whiting/Duchamp/} 
software \citep{Whiting2012}, which has been developed to
automatically detect emission in three-dimensional data. {\sc duchamp}
is specifically designed for the case of a small number of signal
voxels within a large amount of noise and is perfectly suited to the
wide bandwidth data-cubes produced by the Mopra spectrometer.

\begin{figure}
  \centering
  \includegraphics[width=8.3cm, angle=0, trim=0 0 0 0]{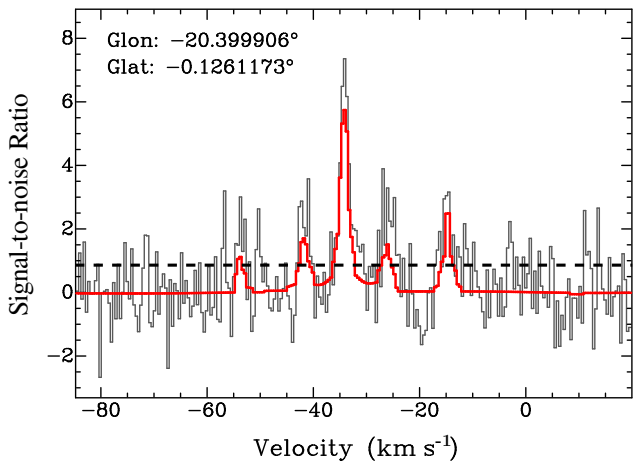}
  \caption{Example signal-to-noise spectrum (grey line) extracted from
    a single pixel with the wavelet-reconstructed spectrum overlaid
    (red-line). {\sc Duchamp} does an excellent job of suppressing the
    noise allowing the use of a low threshold: 0.8$\sigma$ in this
    case.}
  \label{fig:example_recon}
\end{figure}


\subsection{Overview of the source finding procedure}
{\sc duchamp} searches a cube by applying a single threshold, either a flux
value or a signal-to-noise ratio, for the whole dataset, and so is
thus best applied to data with uniform noise. To compensate for the
variable noise in the $\nhthree$ data, the HOPS 
emission finding procedure implements a two-pass solution. Firstly {\sc
  duchamp} is run on a smoothed version of the entire cube down to a
global 3$\sigma$ cutoff level. A mask data-cube is produced which is
used to blank the emission in the original data. Some of this `emission'
identified in the first pass may in fact be regions of high RMS noise
and conversely some real low-level emission may have escaped
detection. However, this 
likely does not matter as the goal of this step is to create a cube
containing {\it mostly} line-free noise. For every spatial pixel in
the blanked cube the standard deviation ($\sigma_{T_{\rm mb}}$) is
estimated along the spectral axis to create a map of background noise
temperature like that shown in Figure~\ref{fig:noise_full}. In
practice we calculate the median absolute deviation from the median
(MADFM), which for a dataset X = x$_1$, x$_2$
\ldots\,x$_i$\ldots\,x$_n$ is given by:  
\begin{equation}\label{eqn:madfm}
  {\rm \sigma = K.median\,(|x_i - median(X)|)},
\end{equation}
i.e., the median of the deviations from the median value.
In Equation~\ref{eqn:madfm}, K is a constant scale factor which depends
on the distribution. For normally distributed data
K\,$\approx$\,1.48. The MADFM statistic is largely
robust to the presence of a few channels much brighter than the
noise. This is especially important when estimating the noise in
spectra containing very broad lines, such as those observed towards
the Galactic centre. 

Each spectral plane of the original input cube is then divided by the
noise map to make a signal-to-noise cube with homogeneous noise
properties. {\sc duchamp} offers the option of reconstructing data-cubes
using the {\it \`a trous} wavelet method prior to running the
finder. A thorough description of the procedure may be found in
\citet{starckmurtagh1994}. The reconstruction is very effective at
suppressing noise in the cube, allowing the user to search reliably to
fainter levels and reducing the number of spurious detections. We
chose to reconstruct the $\nhthree$ data in 3D-mode, meaning that the
wavelet filters can distinguish between narrow noise spikes confined
to single channels and small regions of emission spanning $l$-$b$-$v$
space. Figure~\ref{fig:example_recon} shows an example
$\nhthree$\,(1,1) spectrum with the reconstructed version
overplotted. A second pass of {\sc duchamp} is run on the
reconstructed cube to construct the final list of emission sources. A
detailed description of the {\sc duchamp} inputs is presented in
Appendix~\ref{sec:app_duchamp}.  


\subsection{Cloud measurements}\label{sec:cloud_measure}
The properties of $\nhthree$ clouds were measured directly from the
$\nhthree$ data cubes using the 3D voxel-based emission masks produced by
the source finder. Two dimensional pixel-masks were made by collapsing the
3D masks along the velocity axis. 


\subsubsection{Position, velocity and angular size}
Three position and velocity measurements were made on each cloud: {\it
  centroid}, {\it weighted} and {\it peak}. The centroid
position ($l_{\rm c},\,b_{\rm c},\,v_{\rm c}$) was measured directly
from the 3D voxel mask and corresponds to the geometric centre of the
cloud in $l$-$b$-$v$ space. A brightness weighted position ($l_{\rm
  w},\,b_{\rm w},\,v_{\rm w}$) was measured by weighting each voxel
coordinate with its brightness temperature value $T_{\rm mb}$ according to:
\begin{equation}
n_{\rm w} = \left. \displaystyle\sum\limits_{i=1}^n\,n_i~T_{{\rm mb},i}
\middle/
\displaystyle\sum\limits_{i=1}^n T_{{\rm mb},i}, \right.
\end{equation}
where $n$ is the coordinate axis to be measured. The peak
position ($l_{\rm p},\,b_{\rm p},\,v_{\rm p}$) is the coordinate of
the brightest {\it voxel} in the 3D cloud. Noise spikes can introduce
significant errors in the weighted and peak measurements, especially
for weak detections, so we performed the coordinate measurements on cubes
spatially smoothed to a resolution of 2.5~arcmin and spectrally smoothed
using a hanning window five channels wide. Sources with simple morphologies
have similar positions and velocities in all three measurements
compared to extended sources in which the measurements differ.


\subsubsection{Angular size and area}
The angular size of the clouds was quantified in two ways. Firstly,
the solid angle $\Omega$ subtended by the cloud was measured from the
2D pixel mask. The equivalent angular radius $r_c$ is then defined as the
radius of a circular source which would subtend an equivalent solid
angle on the sky. Secondly, the brightness-weighted radius $r_{\rm w}$ was
calculated with respect to the ($l_{\rm w},\,b_{\rm w}$) position via:
\begin{equation}
r_{\rm w} = \left. \displaystyle\sum\limits_{i=1}^n\,r_i~T_{{\rm mb},i}
\middle/
\displaystyle\sum\limits_{i=1}^n T_{{\rm mb},i}, \right.
\end{equation} 
where $r_i$ is the distance between the brightness-weighted centre and
the $i^{\rm th}$ pixel. For a high
signal-to-noise ratio unresolved source this is directly equivalent to the
Gaussian FWHM/2.


\begin{figure*}
  \centering
  \includegraphics[width=17.5cm, angle=0, trim=0 0 0 0]{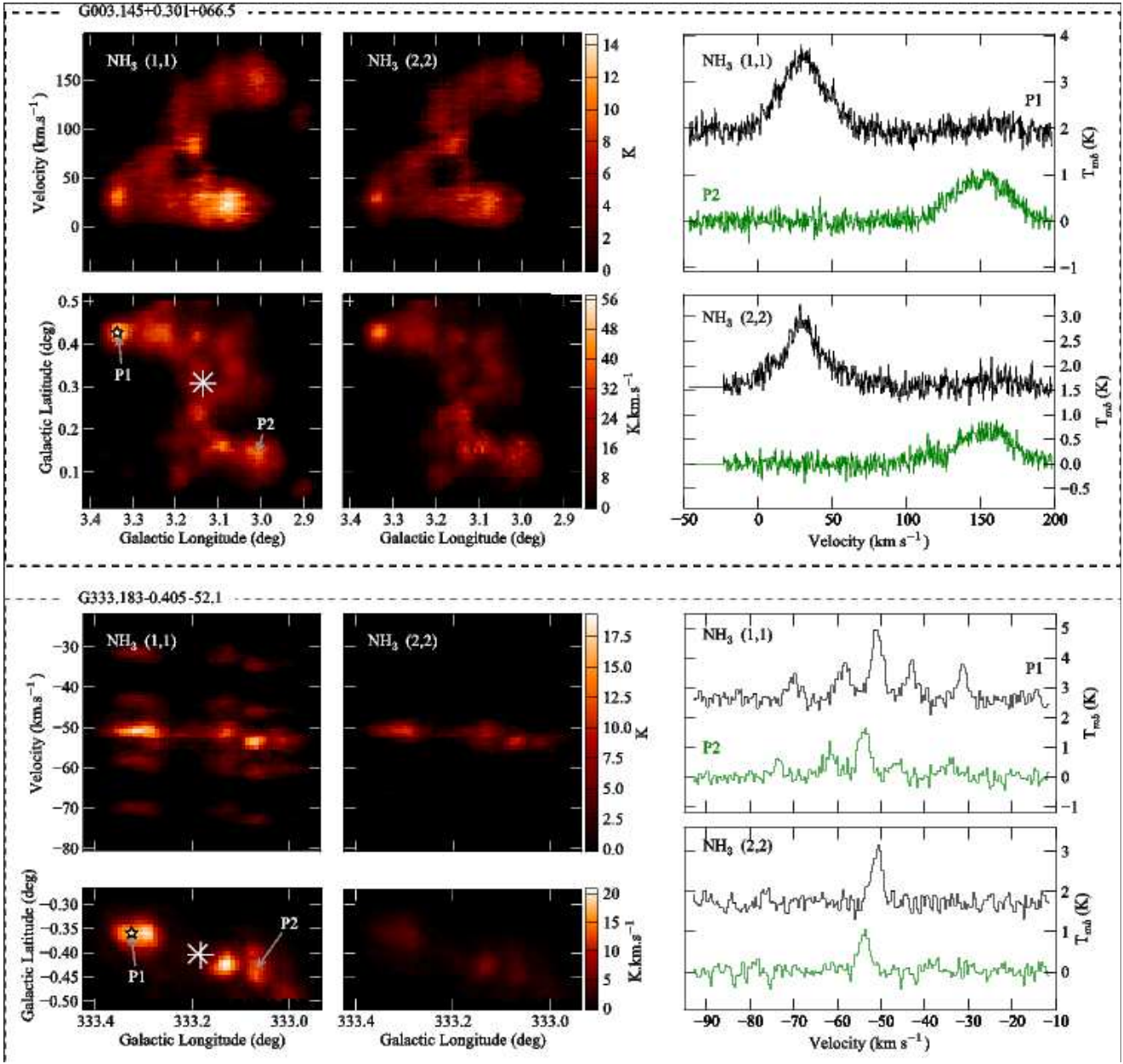}
  \caption{Examples of two bright clouds in HOPS. For each cloud we
    show the \nhthree\,(1,1) and (2,2) integrated intensity maps
    (bottom-left and bottom-middle panels, respectively), summed
    longitude velocity ($l$-$v$) diagrams (top-left and top-middle
    panels) and sample spectra from two local peak positions (right
    panels). The $l$-$v$ diagrams were made by summing over the
    Galactic $b$ axis of the cubes. The position of the brightest
    voxel is marked by a star symbol, the geometric centre by a `$+$'
    and the brightness-weighted centre by a `$\times$'. See
    section~\ref{sec:example_clumps} for details.}  
  \label{fig:example_clumps}
\end{figure*}

\subsubsection{Integrated intensity and brightness temperature}
The integrated intensity $\int T_{\rm mb}\,dv$ of a cloud is the summed
brightness of the emitting voxels (in K) times the velocity-width
$\Delta v$ of a channel (in $\kms$). We have an excellent measurement
of the average RMS  
noise over the cloud (see Figure~\ref{fig:noise_full}), so the
uncertainty on $\int T_{\rm mb}\,dv$ is given by:
\begin{equation}
  \delta\left(\int T_{\rm mb}\,dv\right)~=~\langle RMS\rangle \Delta v\,\sqrt{N},
\end{equation}
where $N$ is the number of independent spatial pixels subtended
by the cloud. Systematic fluctuations in the spectral baselines may also
contribute to the error, but these are likely insignificant compared
to the HOPS sensitivity. 

The peak brightness temperature $T_{\rm mb}$ is simply the value of
the brightest voxel in the cloud.


\subsection{Example clouds}\label{sec:example_clumps}
Forty-four percent of \nhthree\,(1,1) clouds in
HOPS are unresolved ($r_{\rm
  w}<1$~arcmin). Figure~\ref{fig:example_clumps} presents examples of
two bright and extended clouds. Both clouds show evidence 
of sub-structure in position and velocity, but are merged into a
single object by the emission finder. G003.145+0301+066.5 (named
for $l_{\rm c}$+$b_{\rm c}$+$v_{\rm c}$), also known as Bania's Clump
2, \citep{Bania1977}, is interesting as it is one of the few clouds
outside of the CMZ to exhibit very broad line profiles ($\delta
v>50\,\kms$). The emission is largely confined to two lobes separated
in $b$ and $v$, but with components which merge at 
$v\approx75$\,\kms. Physical conditions in G003.145+0301+066.5 are
comparable to the CMZ, as the median brightness ratio
$T_{\rm mb,\,(1,1)}/T_{\rm mb,\,(2,2)}=1.2$ compared to 1.6 for clouds
in the outer Galaxy. 

G333.183$-$0.405$-$52 is a well known giant 
molecular cloud containing H{\scriptsize
  II} regions, hot molecular cores and outflow sources (see
\citealt{lo2009} and references therein). In contrast to 
Bania's Clump 2, the line profiles are well-defined. \nhthree\,(1,1)
exhibits the classic spectral shape at both sampled positions: a
central main group of lines with four symmetric satellite groups,
approximately half as bright as the main-group. The spectrum of
\nhthree\,(2,2) also exhibits four groups of satellite lines
(see \citealt{pickett1998}), but they are generally too weak to be
detected in the HOPS data. 

Overlaid on the  $\int T_{\rm mb}dv$ maps are the three position
measurements as described in Section~\ref{sec:cloud_measure}. The peak
voxel is denoted by a star, the geometric centre is marked by a
`$+$' and the brightness-weighted centre by a `$\times$' symbol. Note
that the peak {\it voxel} is not necessarily coincident with the peak
pixel of the  $\int T_{\rm mb}dv$ map. Geometric and brightness-weighted
positions are generally in close agreement, even in clouds with
complex morphology. Indeed, separations of more than a beam-width
between these two positions and the peak voxel position indicates that
multiple clumps of emission exist within a cloud boundary. The
weighted method is the most robust, so we  adopt the weight values
from now on we when referring to longitude, latitude and $\vlsr$ of a
source.
 

\section{The HOPS \nhthree~catalogue}\label{sec:nh3_catalogue}
The emission finder was run on the \nhthree\,(1,1) and (2,2) datasets
independently to produce the final catalogues of clouds. In total 669
\nhthree\,(1,1) and 248 \nhthree\,(2,2) clouds were found in the HOPS
cubes. In the following sections we examine the catalogues
produced by the source finding routines. We aim to quantify the
robustness of the {\scriptsize DUCHAMP} detections and investigate the
distribution of source properties. 


\subsection{$\nhthree$\,(1,1) versus (2,2) catalogues}
The initial \nhthree\,(1,1) catalogue contained 687 detections of
which 18 were flagged as artifacts. Five detections between $-2<l<2$
with $v_{\rm LSR}<-190\,\kms$ are in fact bright, extended clouds
of emission from the $\nhtwo$ transition. The extremely large
linewidths close to the Galactic centre ($50\,\kms < \Delta v <
100\,\kms$) cause  \nhthree\,(2,2) emission to spill over into the
$\nhone$ bandpass. A further thirteen detections are flagged as they
arise from artifacts caused by the baselining routine. Due to the
polynomial used to fit the baseline (see
Appendix~\ref{sec:baselines}), the spectra at the 
edge of the bandpass can rise slightly above the noise cutoff for
several contiguous velocity channels. If this occurs for enough
contiguous spatial pixels to pass the detection threshold the
emission will be classed as a source. However, such sources are easy
to flag by eye. 

{\sc Duchamp} originally found 324 individual clumps within the
\nhthree\,(2,2) data-cube of which 76 are flagged as
artifacts. Eighteen spurious detections are due to a turned-up
bandpass-edge, or an intrusion by a broad \nhthree\,(1,1)
linewing. The J,K\,=\,(1,1) transition is more easily excited than the
(2,2), hence \nhthree\,(2,2) clouds are only considered valid if
associated with (1,1) emission at the same position. Fifty-eight
low level \nhthree\,(2,2) detections ($T_{\rm mb}<5\sigma$,
$n_{\rm pixels}\le12$) were flagged as spurious for this reason. In 
Section~\ref{sec:completeness}, below, we investigate the formal
catalogue completeness and contamination by spurious sources.


\subsection{Catalogue description}
Tables~\ref{tab:clump_catalogue_11} and~\ref{tab:clump_catalogue_22}
present a sample of the catalogue entries for the \nhthree\,(1,1) and (2,2)
data, respectively. The complete catalogue is available in electronic
format on the HOPS
website\footnote{http://www.hops.org.au}. Columns in the tables
are as follows: (1) the cloud name constructed from the centroid
position of the cloud in Galactic longitude, latitude and
velocity; (2) the brightness-weighted Galactic
longitude, (3) latitude  and (4) $v_{\rm LSR}$ measured from a masked
and smoothed version of the original cube. Columns (5), (6) and (7) contain
the Galactic coordinates ($l$\,\&\,$b$) and velocity ($v_{\rm LSR}$)
of the brightest voxel in the cloud. Columns (8) and (9) tabulate the velocity
range over which {\scriptsize DUCHAMP} detects emission within that
clump; column (10) 
contains the angular radius, in arcminutes, of a circular source subtending an
equivalent solid angle on the sky and column (11) the
brightness-weighted radius measured from the integrated intensity
map. Column (12) contains the solid angle subtended by the cloud in
square arcminutes; (13) the number of spatial pixels in the cloud; (14)
the number of emitting voxels in the cloud, and (15) presents the 
total integrated intensity $\int$T$_{\rm mb}dv$. The peak brightness
temperature is recorded in (16) and the local RMS 
noise temperature in (17). Column (18) contains flags noting if the
cloud touches a survey boundary in $b$ (=X), $l$ (=Y) or $v$ (=Z). The
cloud is flagged with an `M' if multiple velocity components were
detected, i.e., if the velocity range is greater than the expected
velocity width of a single $\nhthree$ spectrum. Here we assume a
FWHM for each line-group of $3\,\kms$, which is typical of gas forming
massive stars. The expected velocity range of an \nhthree\,(1,1) spectrum
from an isolated cloud is then $44.6\,\kms$. Clouds which exhibit
significantly different peak and brightness-weighted velocities
($>5\,\kms$) are likely to contain multiple sub-clouds overlapping in
position and are flagged with an `E'. A small number of
clouds detected in the \nhthree\,(1,1) bandpass are in fact broad
line-wings of \nhthree\,(2,2) encroaching on the (1,1) bandpass and are
flagged as artifacts with an `A' in column 18. Similar artifacts in
the (2,2) data are also marked with an `A' in column 18.

\begin{figure}
  \centering
  \includegraphics[width=6.0cm, angle=-90, trim=0 0 0 0]{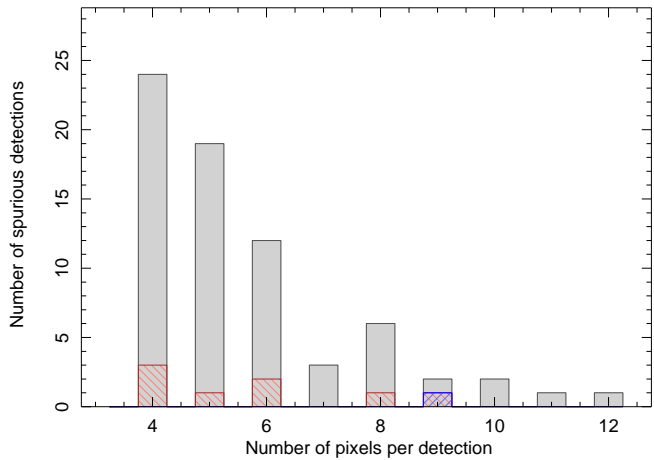}
  \caption{Histogram of the number of spurious sources detected in an
    `empty' \nhthree\,(2,2) data cube (see
    Section~\ref{sec:spurious} for details). The source finder was run  
    at a range of sigma-limits to characterise the number of detections
    as a function of source solid angle. Grey bars\,=\,0.7$\sigma$, 
    red-hatched (left-slanted) bars\,=\,0.8$\sigma$ and blue-hatched
    (right-slanted) bars\,=\,0.9$\sigma$. Based on this plot a brightness 
    cutoff of 0.8$\sigma$ was selected and a minimum of seven spatial pixels
    was required in a valid detection. The artificially low
    $\sigma$-limit is possible because the {\it \`a trous}
    reconstruction effectively suppresses most of the noise.}
  \label{fig:nSpurious}
\end{figure}


\subsection{Catalogue completeness}\label{sec:completeness}


\subsubsection{Spurious sources}\label{sec:spurious}
Although the data have been corrected for large scale inhomogeneities
in the noise properties by fitting polynomial baselines (see
Section~\ref{sec:reduction} and Appendix~\ref{sec:baselines}), higher
order fluctuations are still present and masquerade as real emission
near the sensitivity limits. To constrain the number of spurious
sources as a function of cloud size and sensitivity we ran the source
finder on an `empty' test cube at a range of sigma limits. The test
region was drawn from the \nhthree\,(2,2) dataset in the outer Galaxy
(299.96$^{\circ}$$\,>l>\,$292.54$^{\circ}$, $|b|<\,$0.456$^{\circ}$)
and contains no {\it believable} molecular emission: i.e., when the
source-finder is run at a very low level on both the \nhthree\,(1,1)
and (2,2) data, none of the detected emission is common to the two
datasets. We would expect the J,K\,=\,(1,1) and (2,2) transitions to
arise in the same gas and the \nhthree\,(2,2) line to be weaker,
hence, the detected (2,2) emission without corresponding (1,1)
emission is likely spurious. 

\label{sub:src_completeness}
\begin{figure}
  \centering
  \includegraphics[width=9.0cm, angle=0, trim=0 0 0 0]{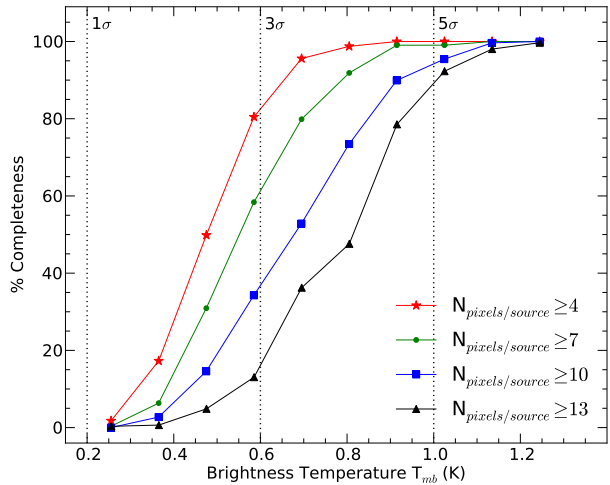}
  \caption{This figure shows the percentage completeness as a
    function of peak brightness temperature. The four curves
    illustrate the effect of varying the minimum number of spatial pixels
    required in a detection. } 
  \label{fig:src_completeness}
\end{figure}

Figure~\ref{fig:nSpurious} shows the number of spurious sources
detected as a function of pixel count and sensitivity limit.
Decreasing both the sensitivity and pixel limits has the effect of
increasing the number of spurious sources found. In particular,
dropping the sigma-limit from 0.8$\sigma$ to 0.7$\sigma$ results in
an sharp increase from eight to seventy spurious detections. The 0.7$\sigma$
results also show that decreasing the minimum size of the clouds
significantly increases the number of spurious detections. Based on these
results we chose a sensitivity limit of 0.8$\sigma$ when source
finding and required a minimum of seven spatial pixels in a valid
detection. 

Data in the test cube is of particularly high quality compared to
the remainder of the survey and may not be wholly representative of all
HOPS data (see Figure~\ref{fig:noise_full}). Fifty-eight spurious
$\nhthree\,(2,2)$ sources (i.e., without associated $\nhthree\,(1,1)$)
were detected in the full survey, implying that significant numbers of
spurious sources exist in the catalogues. However, most of these
detections have signal-to-noise ratios below five and subtend less
than twelve pixels on the sky. More conservative constraints may be
imposed when defining a high-reliability catalogue, e.g., a cutoff of
13 pixels per source would provide a reasonable `safety margin' for
data where the noise properties are worse than in the test
region. Applying such a cutoff to the $\nhthree\,(1,1)$ and (2,2)
catalogues yields source counts of 523 and 198, respectively - a
reduction of $\sim20$ percent.


\subsubsection{Completeness curves}
We have estimated the completeness limits of the catalogue by
injecting artificial point sources into the \nhthree\,(1,1)
data and attempting to recover them using our {\scriptsize DUCHAMP}
procedure. The region of Galactic plane from $l$\,=\,292.4$^{\circ}$ to
$l$\,=\,297.7$^{\circ}$ is devoid of significant emission and was used
as an input. During the experiment the FWHM Gaussian linewidth of the
injected sources was set uniformly to 3\,\kms, typical of high-mass
star-forming clouds. The distribution of peak brightness
temperatures ranged from 0.2\,K to 1.3\,K, spanning the expected
sensitivity limits. One-hundred 3D Gaussian sources were inserted into
the cube at random, non-overlapping 
positions, and the emission finding procedure was called to find
them. After thirty iterations the positions of the clouds found by
{\scriptsize DUCHAMP} were matched with the injected positions and the
recovered sources divided into 
brightness bins. Figure~\ref{fig:src_completeness} plots the
percentage of sources recovered as a function of peak
brightness. Apart from the obvious sensitivity limit, the limiting
parameter on the number of clouds found is the number of pixels
allowed within a valid detection. The figure shows  curves for
N$_{\rm pixels}\,\ge\,4$, 7, 10 and 13 pixels with the 1$\sigma$, 3$\sigma$
and 5$\sigma$ sensitivity limits overplotted. When creating the HOPS
$\nhthree$ cloud catalogue we imposed a limit of $\ge7$ pixels per
valid detection, resulting in a 60 percent completeness level at
3$\sigma$ ($T_{\rm mb}=0.6$\,K). The catalogue is 100 percent complete
at the 5$\sigma$ level ($T_{\rm mb}=1.0$\,K).

\begin{figure}
  \centering
  \includegraphics[width=8.3cm, angle=0, trim=0 0 0 0]{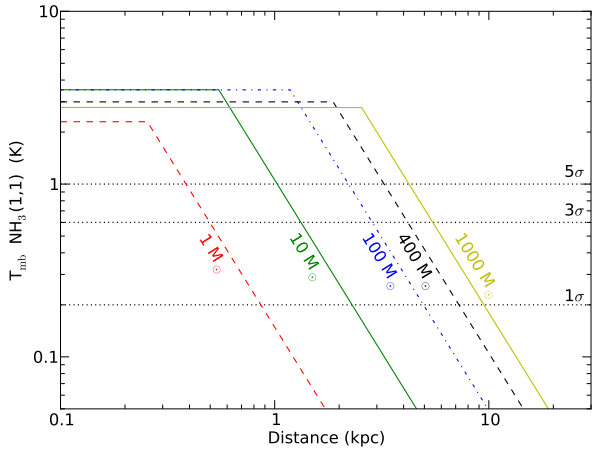}
  \caption{Plot of expected main-line brightness temperature versus
    distance for the five objects in
    Table~\ref{tab:completeness_source_prop}. The one, three and five sigma
    sensitivity limits are plotted as horizontal dashed lines. For the
    parameters outlined in Table~\ref{tab:completeness_source_prop} we
    should detect a 10\,$\msun$ cloud out to a distance of
    $\sim$\,1.0\,kpc, and a 400\,$\msun$ cloud out to 3.2\,kpc at the
    $5\sigma$ level. The turnover occurs as a result of beam dilution:
    once the source is far enough away it no longer fills the main beam.}
  \label{fig:example_sensitivity}
\end{figure}


\subsubsection{Completeness illustration}\label{sec:completeness_illustrated}

\begin{table} 
  \centering
  \caption{Properties of the objects used to illustrate the survey
    completeness. The five objects are taken to be representative of an
    isolated low-mass core and low, intermediate and two high-mass star
    formation regions. A constant $\nhthree$/H$_2$ abundance of
    10$^{-8}$ was assumed. $T^{11}_{\rm ex}$, $T^{11}_{\rm mb}$ and
    $\tau$ were calculated using {\scriptsize RADEX}, as described in
    Section~\ref{sec:completeness_illustrated}.}
\begin{small}
\begin{tabular}{c@{~~~}c@{~~~}c@{~~~}c@{~~~}c@{~~~}c@{~~~}c@{~~~}c@{~~~}}
\hline \hline 
Mass            & Radius     & Density  & $T_{\rm K}$ & $\Delta V^{11}$ & $T^{11}_{\rm ex}$ & $T^{11}_{\rm mb}$ & $\tau$ \\
(M$_\odot$)      & (pc)       & ($\cm3$) & (K)   &  ($\kms$)     & (K)             & (K)             &         \\ \hline
1               & 0.07       & 10$^4$    & 10   & 0.3            & 6.0             &  2.3            & 1.2      \\
10              & 0.16       & 10$^4$    & 10   & 0.3            & 6.6             &  3.5            & 2.3      \\
100             & 0.34       & 10$^4$    & 20   & 1.0            & 8.1             &  3.5            & 1.1      \\
400             & 0.55       & 10$^4$    & 20   & 2.0            & 7.8             &  3.0            & 0.9      \\
1000            & 0.74       & 10$^4$    & 20   & 3.0            & 7.7             &  2.8            & 0.8      \\ 
\hline
\end{tabular}
\end{small}
\label{tab:completeness_source_prop}
\end{table}

To illustrate the completeness limits, we calculated the expected
$\nhone$ brightness temperature as a function of distance for five
objects taken to be representative 
of an isolated low-mass core and low, intermediate and two high-mass
star-forming regions. Starting with spherical clouds of a given mass
and uniform density, we calculated the expected physical radius, projected
angular size and $\nhthree$ column density (from the H$_2$ column
density assuming an H$_2$/$\nhthree$ abundance). The radiation
temperature $T_{\rm R}$, excitation temperature $T_{\rm ex}$ and
optical depth $\tau$ of the $\nhone$ line was then calculated using the
{\scriptsize 
  RADEX}\footnote{http://www.sron.rug.nl/\~vdtak/radex/index.shtml}
radiative transfer package \citep{vanDerTak2007}. {\scriptsize RADEX}
is a one-dimensional non-LTE code that takes as input the kinetic
temperature $T_{\rm kin}$, H$_2$ density, $\nhthree$ column density and
$\nhthree$ linewidth. For each molecular transition it returns
values for $T_{\rm R}$, $T_{\rm ex}$ and $\tau$. The measured main-beam
brightness temperature $T_{\rm mb}$ was calculated from $T_{\rm R}$ by
applying the beam filling factor as a function of source distance.

Figure~\ref{fig:example_sensitivity} shows plots of expected $\nhone$
main-line brightness temperature versus distance for five objects with
parameters outlined in Table~\ref{tab:completeness_source_prop}. Taken
at face value we should be able to detect sources of
1\,M$_\odot$, 10\,M$_\odot$, 100\,M$_\odot$, 400\,M$_\odot$ and
1000\,M$_\odot$ out to distances of 0.4\,kpc, 1.0\,kpc, 2.2\,kpc
3.2\,kpc and 4.2\,kpc, respectively, assuming a 5$\sigma$
limit. Dropping to a 3$\sigma$ level allows the detection of
a 1\,$\msun$ at 0.5\,kpc and boosts the other distances by a factor of
1.3. This will not hold in regions like the Galactic centre where the
bright extended emission and large linewidths effectively increase the
detection threshold by orders of magnitude.


\begin{figure}
  \centering
  \includegraphics[width=8.3cm, angle=0, trim=0 0 0 0]{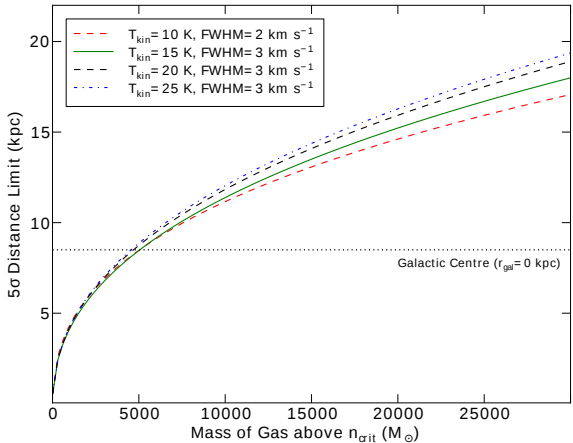}
  \caption{Plot of the 5$\sigma$ detection limit distance versus mass
    of gas above the $\nhthree$ critical density
    ($n_{\rm crit}\approx10^4\,\cmmthree$) for clouds with representative
    excitation conditions. The dotted line shows the distance of the
    Galactic centre at 8.5\,kpc.}
  \label{fig:dist_vs_mass}
\end{figure}

\subsubsection{Comparison with other surveys}
The only comparable single-dish \nhthree\,(1,1) survey conducted in the
southern hemisphere is that of \citet{Hill2010} who used the Parkes
radio-telescope to survey 224 high-mass starforming regions. The
$\sim1$~arcmin
beam FWHM used means their data are four times less beam-diluted than
HOPS and with a median $\sigma_{T_{\rm mb}}=0.05$\,K the data are 4.5
times more sensitive. As a sanity check we cross-matched the HOPS
\nhthree\,(1,1) 
catalogue with the 104 \citet{Hill2010} detections within the HOPS
area. In total seventeen sources are matched within a $2$~arcmin radius,
accounting for the brightest sources in the Hill et
al. catalogue. The unmatched source tend to be weaker and we would not
expect to detect them in HOPS, within the uncertainty imposed by the
unknown beam-dilution factor.


\subsubsection{Galactic coverage}
The globally-averaged volume density of molecular clouds is typically
a few $10^2\,\cmmthree$, and the distribution of molecular gas
across the Galaxy is well-traced by CO emission from various surveys
(e.g., \citealt{dame2001}, \citealt{jackson2006}). The fraction of gas
in these clouds at densities of $10^3-10^4\,\cmmthree$, from which
$\nhthree$ emission can be detected, is usually small -- of order a few
percent. However, it is in the high density gas that star formation
occurs. Recent studies suggest that a critical factor controlling the
star formation rate within a molecular cloud is the fraction of gas
above this density threshold \citep{Lada2012}. A census of this
dense gas fraction is clearly very important. As a large, blind survey
tracing gas at high critical density, HOPS is potentially a powerful
tool for this purpose. 

We have used {\scriptsize RADEX} to estimate how the 5$\sigma$ $\nhone$ 
detection limit translates to a completeness limit for the mass of gas
above the critical density ($\sim10^4\,\cmmthree$) within molecular 
clouds across the Galaxy. Figure~\ref{fig:dist_vs_mass}
illustrates the 5$\sigma$ distance limit as a function of dense gas
mass for clouds with representative temperatures and linewidths. In
the analysis we assume a fixed $\nhthree$/H$_2$ abundance of
$10^{-8}$. We expect to detect all clouds with $>$5,000$\,\msun$ of
dense gas which lie between the Sun and the Galactic centre. Clouds
with $>$30,000$\,\msun$ of dense gas will be detected at distances
corresponding to the other side of the Galaxy ($d\approx17\,$kpc).
However, the distribution of molecular gas in the Galaxy is not
uniform -- one-third is concentrated within 3\,kpc of the Galactic
centre (see the model by \citealt{Pohl2008}). In practice this means
that over 90 percent of clouds which contain $20,000\,\msun$ of dense
gas will be detected in the HOPS survey area. The HOPS survey samples
approximately 66 percent of the volume of the Galaxy where molecular
clouds are formed.

\begin{figure*}
  \centering
  \includegraphics[width=17.7cm, angle=0, trim=0 0 0 0]{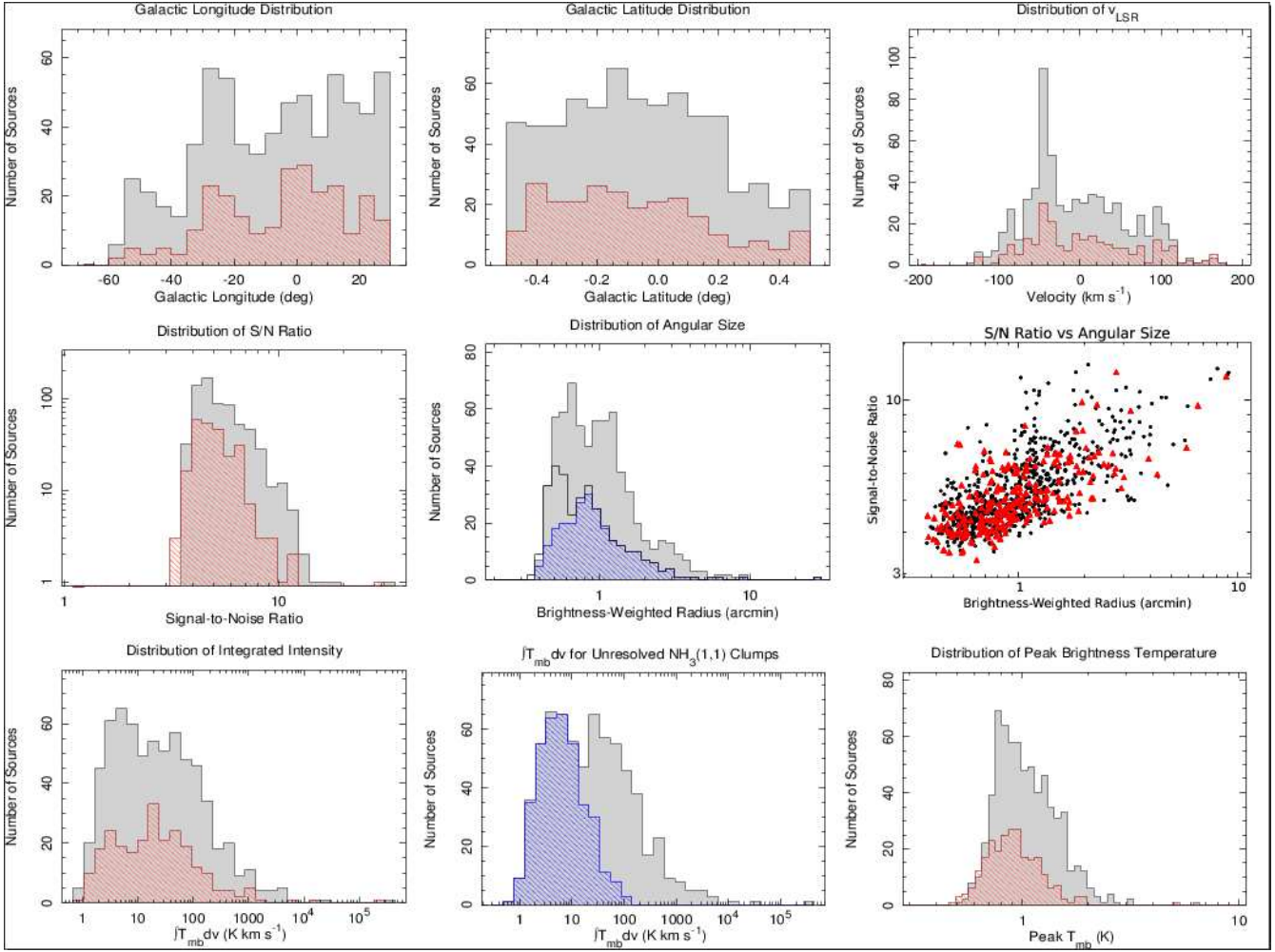}
  \caption{Properties of the clouds in the \nhthree\,(1,1) and (2,2)
    catalogues. The \nhthree\,(1,1) catalogue is represented by grey
    histograms and black symbols, while the \nhthree\,(2,2) catalogue
    is represented by red symbols and hatched histograms. In the
    lower-middle plot the blue hatched histogram is the subset of 
    unresolved \nhthree\,(1,1) clouds. See
    Section~\ref{sec:clump_props} for details.}
  \label{fig:duch_hists}
\end{figure*}


\section{Ensemble cloud properties}\label{sec:clump_props}
In the following sections we investigate the bulk properties of the
clouds in the catalogue.


\subsection{Measured properties}
Figure~\ref{fig:duch_hists} presents selected plots of the measured
properties for all \nhthree\,(1,1) and (2,2) clouds.


\subsubsection{Galactic longitude and latitude}\label{sec:longlat}
The distribution of detections as a function of Galactic longitude is
shown in the top-left panel of Figure~\ref{fig:duch_hists}. Between
$0^\circ<l<30^\circ$ the number of detections is relatively flat. The
CMZ region, which dominated the distribution of integrated intensity
in Figure~\ref{fig:Tbdv_vs_lb}, is detected as a single cloud, while
small peaks in the distribution (e.g., at $l\approx13.5^\circ$ and
$l\approx27.5^\circ$) are due to large star-forming complexes visible
in Figure~\ref{fig:s2n_full}. The longitude range spanning
$-30^\circ<l<-20^\circ$ contains the bright G333.0$-$0.4 high-mass
star-forming cloud and the `Nessie' filament, giving rise to the peak
at $l\approx-25$. There is also a small peak in the number of
detections at $l=-53\degree$, corresponding to the G305 star-forming
complex and a tangent point in the Scutum-Centarus spiral arm. The
longitude distributions of the \nhthree\,(1,1) and (2,2) catalogues
show similar trends, with the \nhthree\,(2,2) catalogue having half as
many sources per bin, as expected from the ratio of total number of
sources in each catalogue. There are no detections in the
$-60^\circ<l<-70^\circ$ range. 

The Galactic latitude source distribution is shown in the top-centre
panel of Figure~\ref{fig:duch_hists} and is significantly biased
towards negative latitudes. 
At $b>0.2^\circ$ the \nhthree\,(1,1) source count falls from less than
30 to approximately 20 sources per 4~arcmin bin. In contrast, no drop is
seen at negative latitudes ($b<-0.2^\circ$), instead the source counts
remain close to 45 per bin. As before both the \nhthree\,(1,1) and
(2,2) distributions exhibit similar shapes. 


\subsubsection{$\vlsr$}\label{sec:vlsr}
The distribution of brightness-weighted $\vlsr$ over the
$\pm\,200\,\kms$ bandpass is shown in the top-right panel of
Figure~\ref{fig:duch_hists}. Similar trends are seen between the
\nhthree\,(1,1) and (2,2) catalogues. The histogram is dominated by
the sharp peak at $\vlsr\approx-45\,\kms$. Upon examination of the
full $l$-$v$ plot for the survey we see that that this feature is
attributable to the sum of velocity components from the southern
Galactic plane between $-70\degree<l<-20\degree$. Discounting the
peak, the shape of the distribution is symmetric, with the majority of
sources falling between $-100\,\kms<\vlsr<100\,\kms$. The
\nhthree\,(1,1) source count beyond  $|\vlsr|>100\,\kms$ drops off
rapidly from $\sim30$ to $<5$. 


\subsubsection{Sensitivity and angular size}\label{sec:sens_size}
The mid-left panel of Figure~\ref{fig:duch_hists} shows the
distributions of signal-to-noise ratio (S/N\,=\,$T_{\rm
  mb}/\sigma_{T_{\rm mb}}$) for the \nhthree\,(1,1) and (2,2)
clouds. In practice the brightness and size limits we imposed on the
source finder mean that we are sensitive to clouds with S/N$\geq3$,
although the majority of clouds fall below S/N$\approx20$. A typical
isolated cloud detected in the $\nhone$ dataset has a FWHM linewidth
across the main group of $2\,\kms$, equivalent to five spectral
channels. Integrating the data across five channels would increase the
signal-to-noise by $\sqrt{5}$ i.e. a $3\sigma$ detection would
become a $6.7\sigma$ `integrated' detection.

The middle panel shows the distribution of brightness-weighted radii
$r_w$ measured from the clouds. On average the \nhthree\,(2,2) clouds
are smaller than the (1,1) clouds. Interestingly, the \nhthree\,(2,2)
histogram exhibits a single peak, while the \nhthree\,(1,1)
distribution is split into two peaks of approximately resolved and
unresolved sources.  We attribute the excess of unresolved detections
to a population of spurious sources similar to the 58 low-level
detections flagged out of the \nhthree\,(2,2) catalogue. The thick
black histogram illustrates the initial \nhthree\,(2,2) detections
{\it including 
  spurious sources}, flagged for having no associated \nhthree\,(1,1)
emission. The flagged sources are clearly responsible for a second
peak corresponding to an excess of unresolved sources.

The mid-right panel of
Figure~\ref{fig:duch_hists} illustrates the correlation between radius
and signal-to-noise ratio. Forty-four percent of the detections in the
survey are are unresolved ($r_{\rm w}<1$~arcmin) and the majority of
these sources have S/N$\le6$.

\begin{figure*}
  \centering
  \includegraphics[width=17.6cm, angle=0, trim=0 0 0 0]{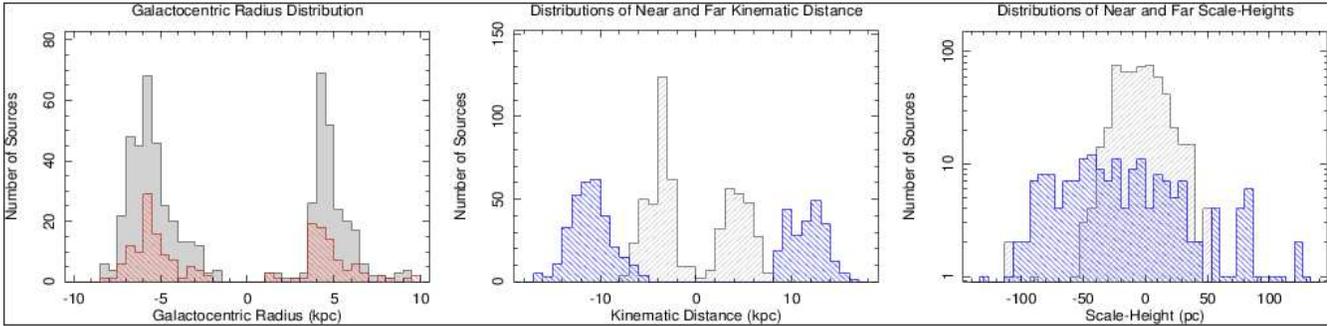}
  \caption{{\it Left panel:} Distributions of galactocentric
    radii for the \nhthree\,(1,1) (solid histogram) and (2,2)
    (hatched histogram) clouds. The two peaks correspond to the 5\,kpc
    Galactic ring at positive and negative $l$. {\it Middle panel:}
    Distributions of near (left-slanted hatch) and far (right-slanted
    hatch) kinematic distances for the \nhthree\,(1,1) clouds
    calculated from the brightness-weighted $v_{\rm LSR}$. {\it Right
      panel:} Near and far projected Galactic scale-heights for the
    \nhthree\,(1,1) clouds. Note: In the first two panels negative
    values on the x-axis signify clouds in the southern Galactic
    plane (i.e., $l<0$). Distances were not determined for the sources
    with $|l|<5\degree$.}  
  \label{fig:dist_hists}
\end{figure*}


\subsubsection{Integrated and peak intensity}\label{sub:integ_int}
The lower-left panel of Figure~\ref{fig:duch_hists} shows the
distribution of integrated intensities for the two catalogues. The
shapes of the J,K\,=(1,1) and (2,2) distributions are similar,
although the \nhthree\,(2,2) integrated intensities are weaker on
average. Again the \nhthree\,(1,1) distribution exhibits a twin-peaked
profile, which can be accounted for by the division into resolved and
unresolved populations. This is demonstrated in the lower-middle panel, which
plots the $\int T_{\rm mb}dv$ distribution for unresolved
\nhthree\,(1,1) clouds overlaid on the total distribution. The
unresolved subset clearly corresponds to the less intense peak, which
likely contains significant numbers of spurious sources.

The difference in the integrated intensities is in part due to the
higher average brightness temperature of the \nhthree\,(1,1) line, as
shown in the bottom-right panel of Figure~\ref{fig:duch_hists}. The
\nhthree\,(1,1) and (2,2) distributions exhibit similar shapes,
although the median $T_{\rm mb}$ for the J,K\,=(1,1) transition is
0.99\,K compared to 0.87\,K for (2,2).


\subsection{Kinematic distance and Galactic structure}\label{sec:gal_struct}
The Galactic disk has a well measured rotation curve which can be used
to solve for an approximate `kinematic' distance given a line-of-site
velocity towards a particular Galactic longitude. This method has the
disadvantage that sources located within the radius of the Sun's orbit
have two valid distance solutions, requiring further information to
distinguish between the near and far distances. In addition, local
velocity deviations due to streaming motions in Galactic spiral arms
lead to a distance uncertainty of order $\pm1$\,kpc
(e.g., \citealt{NguyenLuong2011}).

Outside the CMZ the $\nhthree$\,(1,1) spectra of most clouds exhibit
uncomplicated line profiles (i.e., moderate optical depth with only
one or two velocity components) and the measured
$v_{\rm LSR}$ is an accurate representation of the cloud systemic
velocity. Distance estimates made using $\nhthree$ lines are hence
more accurate than those made using optically-thick tracers (e.g.,
CO\,(1\,--\,0)), whose line profiles may be severely distorted. Bright
maser lines (e.g., the 22\,GHz H$_2$O line) have the advantage of
being detectable at further distances, however, individual spectral
components are often offset from the systemic velocity by greater
than 15\,$\kms$ (e.g., \citealt{Caswell2008}).


\subsubsection{Kinematic distance}
Using the rotation curve of
\citet{brand_blitz1993} and the intensity weighted $l$-$b$-$v$
coordinates we have derived the unique galactocentric radius, near and
far kinematic distances, and projected scaleheights for each
$\nhthree$ cloud. The model assumes a distance to the Galactic centre of
8.5\,kpc and a solar velocity of $220\,\kms$. Although recent work, such
as by \citet{Reid2009}, has suggested that the Solar rotational
velocity is higher than the IAU value of 220\,$\kms$, for simplicity
we adopt the rotational velocity of the original model. A more
detailed analysis incorporating the recent developments will be
presented in a future paper. Galactic rotation is known to  
depart from the approximations of rotation curves towards the inner  
Galaxy, as such we omit sources between
$-5\degree<l<5\degree$. Kinematic distance 
estimates were determined for 576 clouds (86\,\%) and are recorded in
the online version of the HOPS catalogue. The left panel of
Figure~\ref{fig:dist_hists} shows the 
distribution of galactocentric  radii for the \nhthree\,(1,1) and
(2,2) clouds, with negative values indicating positions in the
southern Galactic plane ($l<0$). The \nhthree\,(1,1) and
(2,2) distributions are similar and we continue our analysis using the
\nhthree\,(1,1) clouds only, as they are more numerous. The two strong
peaks at approximately $-5.7$\,kpc and $+4.2$\,kpc correspond to the so called
`Galactic ring' feature, a region that is most likely a supposition of
several spiral arms, and which is thought to contain most of the
molecular gas outside of Central Molecular Zone
\citep{Burton1975,ScovilleSolomon1975}. The difference 
in the galactocentric radii of the peaks may be attributed to the
negatively-biased longitude coverage, but it is also consistent with
their origin in an asymmetric structure, such as a spiral arm. 

In the middle panel are plotted histograms of near and far kinematic
distances for the \nhthree\,(1,1) detections. By design the two
distributions have very little overlap, with most near distances
falling between two and eight kpc (inside the tangent circle), and
most far distances between eight and fourteen kpc. Clouds with masses
or kinetic temperatures 
bright enough to be detected beyond the Galactic centre will be
necessarily rare (see Figure~\ref{fig:dist_vs_mass}), implying that
the near distance is favoured for the majority of sources. Support for
choosing the near kinematic distances in at least half of the
$\nhthree$\,(1,1) detections comes from  comparing the catalogue with 
infrared dark clouds (IRDCs). IRDCs are seen as dark extinction
features against the bulk of diffuse Galactic infrared emission and,
due to their nature, are thought to lie at near kinematic distances
\citep{Simon2006b}. We cross-matched the \nhthree\,(1,1) catalogue
with the list of IRDCs compiled by \citet{Simon2006a} from the
8.3\,$\micron$ Midcourse Space Experiment (MSX) images. Fifty
percent of the \nhthree\,(1,1) cloud centres were found to lie within
2~arcmin of an IRDC extinction peak, confirming their association.

We urge caution in considering the scale-height distribution of clouds
in HOPS. The $b\pm0.5\degree$ coverage likely truncates the true 
distribution of molecular gas, as evidenced by the flat emission
profile in Figure~\ref{fig:Tbdv_vs_lb}. Near and far scale-heights are
plotted in the right-panel of Figure~\ref{fig:dist_hists} and have
standard-deviations of 13.1\,pc and 30.9\,pc, respectively. Both
values are within the expected scaleheights of massive star-formation
tracers found in the literature, e.g., \citet{Urquhart2011} measured
an average scaleheight of $29\pm0.5$\,pc for massive young stellar
objects in the Red MSX source survey. 


\subsubsection{Galactic structure}
\begin{figure}
  \centering
  \includegraphics[width=8.2cm, angle=0, trim=0 0 0 0]{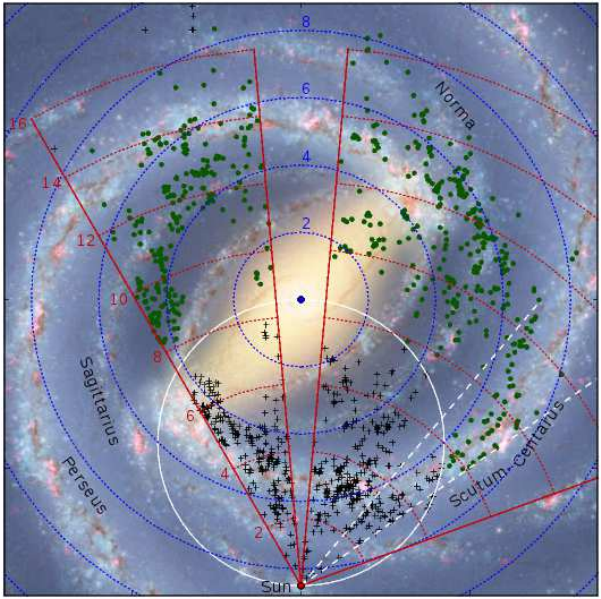}
  \caption{Face-on artistic rendering of the Galaxy showing the
    positions of the clouds found by HOPS at the near and far kinematic
    distances (black crosses and green circles, respectively). This
    plot is modeled after Figure~9 of \citet{Urquhart2011}. Dotted
    lines mark the distance from the Galactic centre (blue) and the
    Sun (red) in kiloparsecs. The four main spiral-arms are labeled in
     black text. The position of the rotational tangent point is
     delineated by a white circle. } 
  \label{fig:gal_face_plot}
\end{figure}
Figure~\ref{fig:gal_face_plot} illustrates the positions of the
\nhthree\,(1,1) clouds on an artists rendering of the Galaxy as viewed
face-on. The image\footnote{http://www.spitzer.caltech.edu/images}
 has been constructed by Robert Hurt of the {\it
  Spitzer} Science Centre in collaboration with Robert Benjamin of the
University of Wisconsin-Whitewater and attempts to include the most
recent information on Galctic structure. The key features are two
`major' spiral-arms (Scutum-Centarus and Perseus), two minor spiral
arms (Norma and Sagittarius) and nested Galactic bars: a long (thin) bar at
an angle of $\sim45\degree$ to the sun-centre line
\citep{Hammersley2000, Benjamin2005, Cabrera-Lavers2008} and a short
(boxy/bulge) bar at an orientation of $20\,-\,30\degree$
\citep{Blitz1991, Weiland1994}. Both near and far kinematic 
distances have been plotted as black crosses and green circles,
respectively. For clouds within 
$-30\degree<l<30\degree$ the near and far distances are widely
separated, with the majority of the far distances lying well beyond
our 3.2\,kpc sensitivity limit for 400\,$\msun$ objects. If we assume
near distances are correct, then the cluster of sources between 
$23\degree<l<30\degree$ and at a distance of 6\,kpc line up with the
intersection of the long-bar and the Scutum-Centarus spiral
arm. \citet{Urquhart2011} also found a large number of high-mass stars
concentrated in this region, while the BUFCRAO Galactic Ring Survey
\citep{roman-duval2009} reported a large amount of molecular material
in approximately the same area. 

Clouds between $-55\degree<l<-40\degree$ (between white dashed lines in
Figure~\ref{fig:gal_face_plot}) correlate well with the Scutum-Centarus
spiral arm when placed at the far distance, and would otherwise fall
between arms. 

A more thorough investigation of the near/far distance ambiguity will
be the subject of a future paper at which time we will be in a
position to comment on Galactic structure with greater authority.



\section{Summary and future work}
We have mapped 100 square degrees of the Galactic plane in the
J,K\,=\,(1,1) and (2,2) transitions of $\nhthree$ at a resolution of
$2$~arcmin using the Mopra radio-telescope. The survey covers the region 
$-70\degree<l<30\degree$, $|b|<0.5\degree$ with a velocity range of
$\pm200\,\kms$. The median sensitivity is
$\sigma_{T_{\rm mb}}=0.20\pm0.06$\,K in each $0.42\,\kms$ spectral
channel. We have developed an automatic emission-finding routine based on 
{\scriptsize DUCHAMP} and used it to find clouds of contiguous
emission in the $l$-$b$-$v$ data cubes. The following are our main findings:
\begin{enumerate}
  \item We have detected $\nhthree$ emission across the Galactic
    longitude range between $-60\degree<l<30\degree$, however, the
    outer Galaxy between $-70\degree<l<-60\degree$ is devoid of
    emission above our sensitivity limit. The Central Molecular
    Zone (CMZ, $|l|<2\degree$) contains 80.6 percent of the
    \nhthree\,(1,1) integrated intensity, which appears as a single
    giant cloud at 2~arcmin resolution. Most of the remaining gas
    is concentrated within $2\degree<|l|<30\degree$. 
    Within the CMZ the \nhthree\,(1,1) integrated intensity
    peaks near the Galactic midplane, but the distribution with $b$ is
    largely flat outside the CMZ.
  \item Using our emission-finding procedure we detected 669
    \nhthree\,(1,1) clouds and 248 \nhthree\,(2,2) clouds, and
    measured their basic properties. Forty-four percent of the
    \nhthree\,(1,1) clouds are unresolved in the 2~arcmin Mopra
    beam. The full HOPS $\nhthree$\,(1,1) catalogue likely contains
    significant numbers of unresolved spurious sources below
    $5\sigma$. However, a high-reliability catalogue may be constructed by
    restricting the number of pixels required in a detection to 13, reducing
    the source-count by 20 percent.
  \item The $\nhthree$ catalogue contains clouds detected down to a
    3$\sigma$ level (0.6\,K) where it is 60 percent complete. The
    catalogue is $\sim100$ percent complete at the 5$\sigma$ level
    (1.0\,K). As an illustration of the sensitivity, we would detect a
    typical 400\,M$_{\odot}$ \nhthree\,(1,1) cloud (T$_{\rm kin}=20$\,K,
    $n_{\rm H_2}=10^4\cmmthree$) out to a distance of 3.2\,kpc at the
    5$\sigma$ level. Similar clouds containing 5,000\,$\msun$ and
    30,000\,$\msun$ would be detected at the Galactic centre and on the
    far side of the Galaxy, respectively.
  \item We have estimated the near and far kinematic distances towards
    the \nhthree\,(1,1) clouds. Given our sensitivity limits the
    majority of clouds likely lie at the near distance. Sources between
    $-55\degree<l<-40\degree$ may lie at either the near or far distance in
    the Scutum-Centarus spiral arm. Further investigation is required
    to distingush between the two possibilities.
\end{enumerate}

Combined with other Galactic plane surveys the HOPS catalogues provide
an invaluable tool for the investigation of Galactic structure and
evolution. The next paper in the series (Longmore et al. {\it in
  prep.}) will present an analysis of the $\nhthree$ spectra, the
physical and evolutionary status of the detected clouds, and the
procedures developed to automate the line fitting. Further papers will
publish the other spectral lines observed, including the J,K\,=\,(3,3)
(6,6) and (9,9) $\nhthree$ transitions, HC$_3$N\,(3,2) and H69$\alpha$. 


\subsection{Data release}
The $\nhthree$ data and cloud catalogues are available to the
community from the HOPS website, http://www.hops.org.au,
via an automated  cutout server and catalogue files. In
addition to the FITS format data-cubes we have made available the
emission-finder masks, integrated intensity and peak temperature
maps. Interested parties may also download the emission-finding and
baseline-fitting procedures, which have been written in the {\it python}
language.


\section*{Acknowledgements}
The HOPS team would like to thank the anonymous referee whose comments
greatly improved this work. We would also like to thank the dedicated
work of CSIRO Narrabri staff who supported the observations beyond the
call of duty. The University of New South Wales Digital Filter Bank
used for the observations (MOPS) with the Mopra Telescope was provided
with support from the Australian Research Council, CSIRO, The
University of New South Wales, Monash University and The University of
Sydney. The Mopra radio telescope is part of the Australia Telescope
National Facility which is funded by the Commonwealth of Australia for
operation as a National Facility managed by CSIRO. PAJ acknowledges
partial support from Centro de Astrof\'\i sica FONDAP 15010003 and the
GEMINI-CONICYT FUND.

\bibliography{cpurcell_HOPS}
\bsp

\begin{landscape}
  \begin{table}
    \begin{minipage}{235mm}
      \caption{Properties of the \nhthree\,(1,1) clouds identified in HOPS.}
      \begin{small}\label{tab:clump_catalogue_11}
        \begin{tabular}{l@{~~}c@{~~~}c@{}c@{\,}c@{~~~}c@{~~~}c@{~}c@{~\,}c@{~}c@{~~}c@{~~}c@{~~~}c@{~~}c@{~~}c@{~~}c@{~~}c@{~}c@{~~}c@{~~}l@{}l@{}l@{}l@{}l@{}l}
          \hline
          \hline
          & \multicolumn{3}{c}{Weighted} && \multicolumn{3}{c}{Peak} \\
          \cline{2-4}  \cline{6-8}
          Cloud Name$^{*}$ \rule[-0mm]{0pt}{2mm}
          & $l$         & $b$         & $v_{\rm LSR}$ && $l$         & $b$         & $v_{\rm LSR}$ && $v_{\rm min}$ & $v_{\rm max}$ & $r_{\rm c}$ & $r_{\rm w}$ & $\Omega$  & n$_{\rm pix}$  & n$_{\rm vox}$ & $\int T_{\rm mb}dv$     & $T_{\rm peak}$ & $T_{\rm RMS}$ & \multicolumn{5}{@{}l}{Flags$^{\dagger}$}  \\
          \rule[-1.5mm]{0pt}{4mm}
          & ($^{\circ}$) & ($^{\circ}$) & ($\kms$)     && ($^{\circ}$) & ($^{\circ}$) & ($\kms$)     && \multicolumn{2}{c}{($\kms$)}  &($'$)  & ($'$)     & $arcmin^{2}$ &         &       &(K\,$\kms$)  & (K)       & (K)     \\
  (1)     & (2)         & (3)         & (4)         && (5)         & (6)         & (7)          && (8)    & (9)         & (10)         & (11)   & (12)         & (13)    & (14)  & (15)        &(16)      & (17)     & \multicolumn{4}{@{}l}{(18)}  \\
      \hline
${\rm G029.987-0.157+097.3}$ &  29.986 &  -0.156 &  97.3 &&  29.992 &  -0.143 &  98.0 &&  96.3 &  98.0 &  1.2 &  0.8 & 4.75 &    19 &    76 & 7.38 $\pm$ 0.61 & 0.74 & 0.16 & &  &  &  &  \\
${\rm G018.122+0.359+022.4}$ &  18.122 &   0.359 &  22.5 &&  18.125 &   0.357 &  23.0 &&  11.5 &  23.4 &  1.6 &  1.0 & 8.25 &    33 &    83 & 9.75 $\pm$ 0.53 & 0.66 & 0.14 & &  &  &  &  \\
${\rm G008.621-0.346+037.3}$ &   8.622 &  -0.349 &  37.5 &&   8.700 &  -0.401 &  39.2 &&  12.3 &  60.5 &  9.4 &  8.1 & 275.50 &  1102 & 18506 & 2751.12 $\pm$ 10.16 & 2.17 & 0.17 &X&  &  &  &  \\
${\rm G002.507-0.032-02.0}$ &   2.505 &  -0.029 &  -2.4 &&   2.508 &  -0.026 &  -9.0 && -27.7 &  20.9 &  4.0 &  1.9 & 49.25 &   197 &  8250 & 962.50 $\pm$ 7.43 & 1.31 & 0.19 & &  &  &  & E& \\
${\rm G000.860-0.067+045.7}$ &   0.807 &  -0.064 &  47.5 && 359.875 &  -0.084 &  11.9 && -138.5 & 146.6 & 34.5 & 28.7 & 3748.00 & 14992 & 2114971 & 443317.78 $\pm$ 113.19 & 6.40 & 0.18 &X&  &  & M& E& \\
${\rm G359.084+0.108+139.3}$ & 359.078 &   0.109 & 138.9 && 359.042 &   0.141 & 130.8 && 108.7 & 165.4 &  5.4 &  3.1 & 91.00 &   364 & 17181 & 2612.32 $\pm$ 12.95 & 1.56 & 0.23 & &  &  & M& E& \\
${\rm G343.423-0.352-27.3}$ & 343.423 &  -0.351 & -27.2 && 343.425 &  -0.343 & -27.3 && -46.5 &  -7.3 &  3.7 &  2.0 & 42.00 &   168 &   920 & 182.69 $\pm$ 2.99 & 1.54 & 0.23 & &  &  &  &  \\
${\rm G302.008+0.058+198.1}$ & 302.008 &   0.058 & 198.1 && 302.008 &   0.049 & 197.8 && 197.8 & 198.2 &  0.9 &  0.5 & 2.50 &    10 &    13 & 1.10 $\pm$ 0.32 & 0.60 & 0.21 & &  & Z&  &  \\

      \hline
    \end{tabular}
    \end{small}
    \begin{footnotesize}
      $^*$~The cloud is named for the centroid position in $l$-$b$-$v$ space.\\
      $^{\dagger}$~The flags in column (18) have the following
      meanings: X\,=\,the cloud touches a Galactic longitude boundary,
      Y\,=\,the cloud touches a Galactic latitude boundary, Z\,=\,the
      cloud touches a velocity boundary
      ($\pm$\,200\,\kms). M\,=\,multiple velocity components along the
      line of sight, E\,=\,an extended cloud which may contain
      multiple sub-clouds, A\,=\,a cloud identified as artefacts (e.g.,
      invading $\nhthree$\,(2,2) emission or up-turned bandpass-edges). 
    \end{footnotesize}
   \end{minipage}
\end{table}\begin{table}
    \begin{minipage}{215mm}
      \caption{Properties of the \nhthree\,(2,2) clouds identified in HOPS.}
      \begin{small}\label{tab:clump_catalogue_22}
        \begin{tabular}{l@{~~}c@{~~~}c@{}c@{\,}c@{~~~}c@{~~~}c@{~}c@{~\,}c@{~}c@{~~}c@{~~}c@{~~~}c@{~~}c@{~~}c@{~~}c@{~~}c@{~}c@{~~}c@{~~}l@{}l@{}l@{}l@{}l@{}l}
          \hline
          \hline
          & \multicolumn{3}{c}{Weighted} && \multicolumn{3}{c}{Peak} \\
          \cline{2-4}  \cline{6-8}
          Cloud Name$^{*}$ \rule[-0mm]{0pt}{2mm}
          & $l$         & $b$         & $v_{\rm LSR}$ && $l$         & $b$         & $v_{\rm LSR}$ && $v_{\rm min}$ & $v_{\rm max}$ & $r_{\rm c}$ & $r_{\rm w}$ & $\Omega$  & n$_{\rm pix}$  & n$_{\rm vox}$ & $\int T_{\rm mb}dv$     & $T_{\rm peak}$ & $T_{\rm RMS}$ & \multicolumn{5}{@{}l}{Flags$^{\dagger}$}  \\
          \rule[-1.5mm]{0pt}{4mm}
          & ($^{\circ}$) & ($^{\circ}$) & ($\kms$)     && ($^{\circ}$) & ($^{\circ}$) & ($\kms$)     && \multicolumn{2}{c}{($\kms$)}  &($'$)  & ($'$)     & $arcmin^{2}$ &         &       &(K\,$\kms$)  & (K)       & (K)     \\
  (1)     & (2)         & (3)         & (4)         && (5)         & (6)         & (7)          && (8)    & (9)         & (10)         & (11)   & (12)         & (13)    & (14)  & (15)        &(16)      & (17)     & \multicolumn{4}{@{}l}{(18)}  \\
      \hline

      ${\rm G029.995+0.156-84.1}$ &  29.994 &   0.156 & -84.1 &&  29.992 &   0.166 & -39.2 && -84.4 & -83.6 &  0.7 &  0.5 & 1.75 &     7 &    12 & 0.86 $\pm$ 0.27 & 0.71 & 0.18 & &  &  &  & E& \\
${\rm G023.332-0.253+096.4}$ &  23.333 &  -0.253 &  96.4 &&  23.267 &  -0.243 &  60.5 &&  78.0 & 102.3 &  2.0 &  1.5 & 12.25 &    49 &   120 & 16.45 $\pm$ 0.85 & 0.90 & 0.18 & &  &  &  & E& \\
${\rm G018.859-0.477+065.3}$ &  18.860 &  -0.476 &  65.4 &&  18.883 &  -0.476 &  65.6 &&  55.8 &  68.2 &  2.8 &  2.0 & 24.00 &    96 &   677 & 136.29 $\pm$ 2.86 & 1.85 & 0.26 &X&  &  &  &  \\
${\rm G009.125-0.153+043.7}$ &   9.125 &  -0.154 &  43.7 &&   9.117 &  -0.159 &  43.9 &&  43.5 &  43.9 &  0.7 &  0.4 & 1.75 &     7 &    11 & 1.45 $\pm$ 0.27 & 0.80 & 0.19 & &  &  &  &  \\
${\rm G000.892-0.057+050.7}$ &   0.819 &  -0.060 &  49.7 && 359.875 &  -0.084 &  13.2 && -128.7 & 152.6 & 32.6 & 27.5 & 3339.00 & 13356 & 1640704 & 274605.81 $\pm$ 91.23 & 4.96 & 0.17 &X&  &  & M& E& \\
${\rm G338.920-0.483-36.8}$ & 338.922 &  -0.482 & -36.9 && 338.942 &  -0.484 & -37.5 && -37.9 & -35.8 &  1.9 &  1.2 & 11.50 &    46 &   175 & 16.41 $\pm$ 1.04 & 1.17 & 0.18 &X&  &  &  &  \\
${\rm G322.258+0.166-171.2}$ & 322.258 &   0.166 & -171.2 && 322.258 &   0.166 & -171.8 && -171.8 & -170.9 &  0.9 &  0.5 & 2.50 &    10 &    20 & 2.02 $\pm$ 0.31 & 0.67 & 0.16 & &  & Z&  &  \\
${\rm G300.464-0.498-198.0}$ & 300.464 &  -0.497 & -198.0 && 300.450 &  -0.493 & -198.2 && -198.2 & -197.8 &  0.9 &  0.8 & 2.75 &    11 &    21 & 2.40 $\pm$ 0.40 & 1.41 & 0.21 &X&  & Z&  &  \\
      \hline
    \end{tabular}
    \end{small}
    \begin{footnotesize}
      $^*$,\,$^{\dagger}$~Annotation markings have the same meanings as in Table~\ref{tab:clump_catalogue_11}.
    \end{footnotesize}
   \end{minipage}
\end{table}
\end{landscape}

\appendix


\section{Spectral baselines}\label{sec:baselines}

\begin{figure}
  \centering
  \includegraphics[width=8.3cm, angle=0, trim=0 0 0 0]{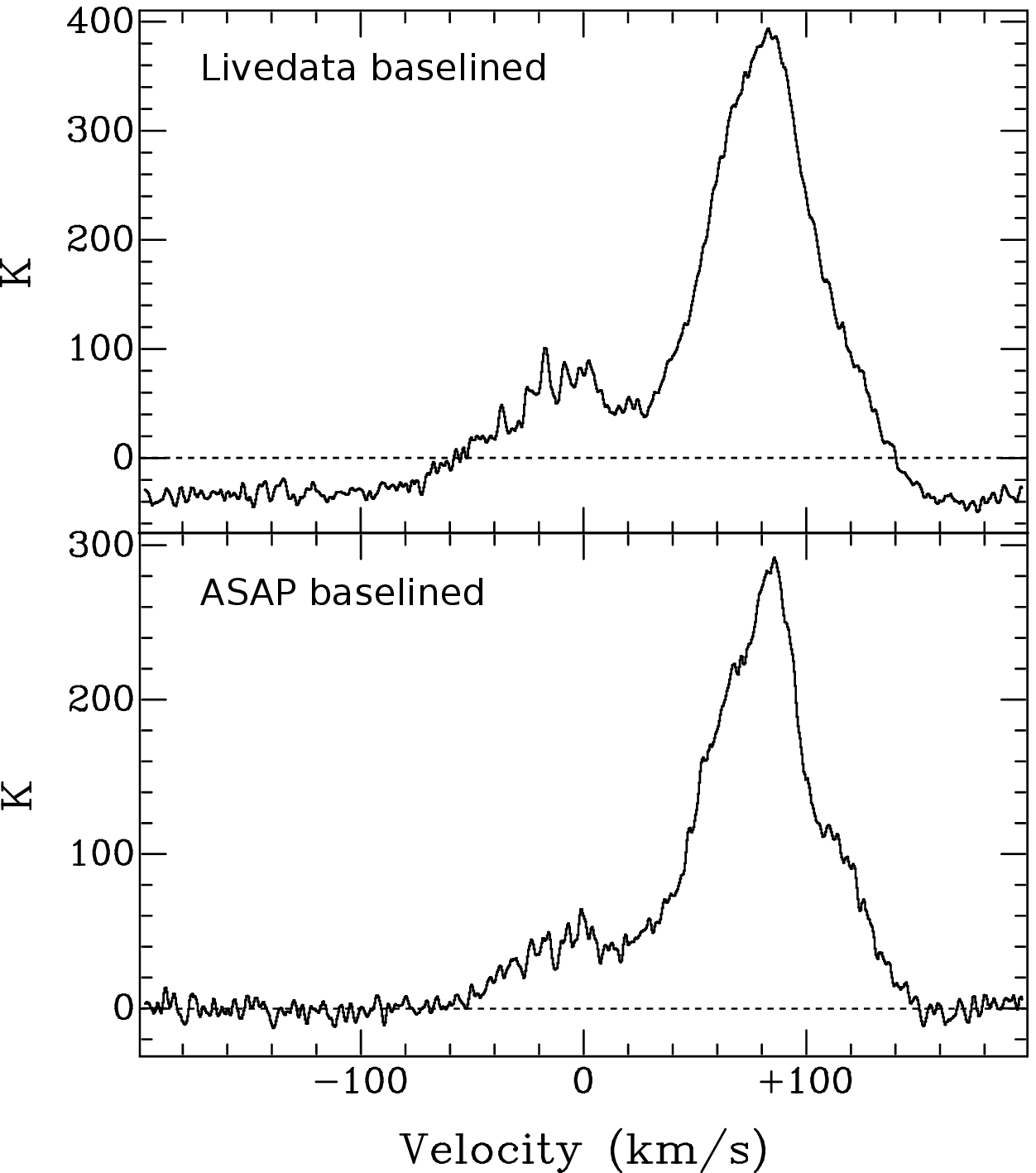}
  \caption{Integrated spectrum measured from an area of
    emission close to the Galactic centre. The spectrum in the top
      panel is drawn from the data baselined using only a 1$^{\rm
      st}$ order polynomial in {\sc livedata}. The spectrum in the
     bottom panel has had further baselines subtracted using the
    ASAP fitting software.} 
  \label{fig:broadline}
\end{figure}

During the initial data reduction, the {\scriptsize LIVEDATA} software
was used to fit and subtract a first order polynomial from the
line free channels of the spectra. In the majority of cases this
was more than sufficient to flatten the baselines, however, we
occasionally found that two types of baseline artifacts remain:
\begin{enumerate}
  \item Ripples and sine-waves caused by rapidly changing weather
    conditions.
  \item Negative bowls adjacent to very broad spectral lines in the
    vicinity of the Galactic centre.
\end{enumerate}
Type (i) artifacts are likely to contaminate the catalogue with
spurious sources where the peaks of the ripples rise above the
noise. To mitigate against this problem we utilised the ATNF
ASAP\footnote{http://svn.atnf.csiro.au/trac/asap} package to fit 
additional polynomials of order 3\,--\,5 to the data. Before performing
the fit, any bright spectral lines were masked off using a prior {\sc
  duchamp}-produced mask. This approach was entirely successful in
eliminating these baseline artifacts.

Broad-line sources with type (ii) artifacts tend to have artificially
clipped boundaries and suppressed flux
densities. Figure~\ref{fig:broadline} shows examples of integrated
spectra measured close to the Galactic centre. The spectrum in the top
panel is extracted from data processed using {\sc
 livedata} and {\sc gridzilla} only.  {\sc livedata} cannot
identify broad spectral features and so fits the wings of
these lines as part of the emission-free baseline, suppressing the
real baseline below zero. In order to zero the baselines we iterated
around the following procedure until no further improvement 
was seen: 
\begin{itemize}
  \item Run the emission finder to produce an emission mask.
  \item Expand the cloud masks by ten percent in $l$-$b$-$v$ .
  \item Mask the emission and fit a first-order polynomial to the
    line-free channels. 
\end{itemize}
The solid line in the bottom panel of Figure~\ref{fig:broadline} shows the
same spectrum after three iterations of fitting. All negative bowls
were successfully eliminated from the cubes.


\section{Running the {\sc Duchamp} emission finder}\label{sec:app_duchamp}
\begin{table*}
\begin{minipage}{174mm}
\centering
\caption{Critical inputs to the {\scriptsize DUCHAMP} emission
  finder (version 1.1.10).}
\label{tab:duchamp_inputs}
\begin{tabular}{lrl}
  \hline
  Input Name     & Value     & Notes\\
  \hline
  flagAtrous     & Yes       & Perform the {\it \`a trous} wavelet reconstruction.\\
  reconDim       & 3         & Reconstruct the data in 3D mode.\\
  scaleMin,\,Max & 1\,--\,0  & Automatically determine which wavelet scales to be included in the reconstructed image.\\
  snrRecon       & 10        & Signal-to-noise ratio required to include a wavelet coefficient at a point in the reconstructed cube.\\
  filterCode     & 1         & Use a B$_3$ spline filter in the reconstruction.\\
  searchType     & Spatial   & Search one channel-map at a time.\\
  threshold      & 0.8       & Search the signal-to-noise cube to a level of 0.8-$\sigma$.\\
  flagGrowth     & False     & Do not `grow' clumps to a lower threshold.\\
  threshSpatial  & 2         & Merge clumps within two spatial pixels.\\
  threshVelocity & 53\,(68)  & Merge clumps within 53 channels (22.3\,\kms), the velocity difference between the central and outer \\
                 &           & satellite groups in the \nhthree\,(1,1) spectrum, plus a 3\,\kms~guard band. For \nhthree\,(2,2) the threshold\\
                 &           & is 68 channels (29.0\,\kms).\\
  minChannels    & 2         & Require a valid detection to span at least two channels (0.85\,\kms).\\
  minPix         & 7         & Require a detection to contain at least seven spatial pixels.\\
  \hline
\end{tabular}
\end{minipage}
\end{table*}

For detailed information on the {\sc duchamp} algorithm and
settings the reader is referred to the {\sc duchamp} web page and
user-guide linked therein. Here we detail how the emission finder
was used with the HOPS $\nhthree$ data.

\begin{figure}
  \centering
  \includegraphics[width=8.2cm, angle=0, trim=0 0 0 0]{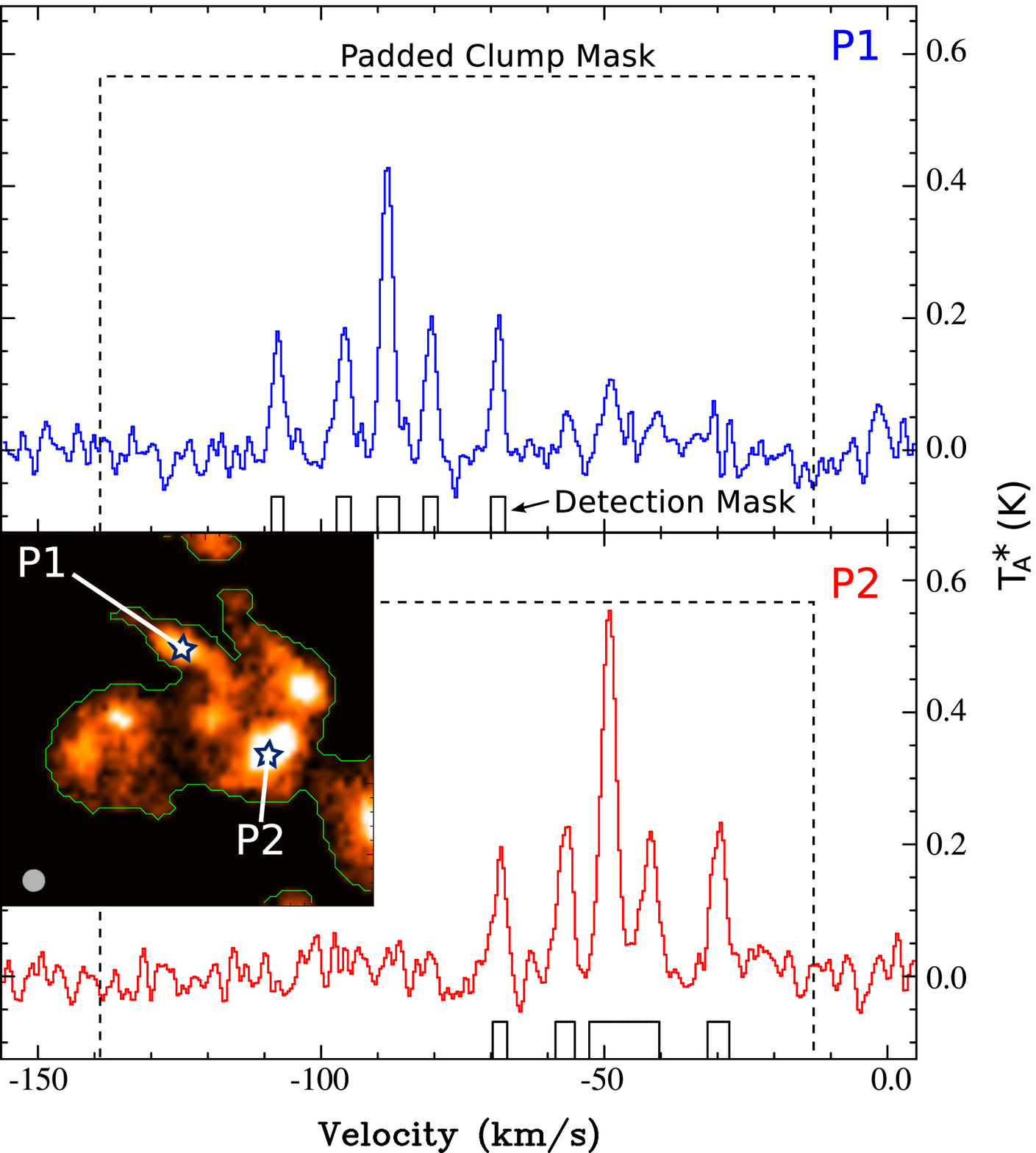}
  \caption{The inset image shows an integrated intensity map of
    $\nhthree$ emission in the G333 cloud. The green contour represents
    the cloud boundary. {\sc duchamp} detects
    each of the \nhthree\,(1,1) spectral line components individually
    (top and bottom panels), however, the sub-clouds overlap both
    spatially and in velocity and are hence catalogued as a single
    merged cloud.} 
  \label{fig:specmask}
\end{figure}
At the outset the reconstructed signal-to-noise cube is searched one
spectral plane at a time down to a global brightness threshold. Voxels
containing emission above the threshold are recorded and grouped with
their immediate neighbours. The list of detections is condensed by
rigorously merging adjacent objects within two spatial pixels 
(1'\,=\,1/2 beam FWHM) or within the velocity
spread of the $\nhthree$ spectrum. The expected velocity limits
correspond to the frequency difference between the main and outer
satellite groups, plus a guard band to account for the width of the
lines. For \nhthree\,(1,1) this $\sim19.7\,\kms$ plus
$3.0\,\kms$ (=\,53 channels) and for \nhthree\,(2,2)
$\sim26.0\,\kms$ plus $3.0\,\kms$ (=\,68 channels). {\sc
  duchamp} makes no attempt to separate contiguous areas of emission into
components in the manner of {\sc clumpfind} \citep{williams1994} or
{\sc fellwalker}\footnote{Part of the {\scriptsize STARLINK} software
  suite, which is available at
  http://starlink.jach.hawaii.edu/starlink}. Adjacent regions of
emission in which 
the outer satellite lines overlap are merged into a single object in 
the resultant catalogue. Figure~\ref{fig:specmask} shows an example of
an emitting region containing several distinct velocity components which
are counted as a single cloud by {\sc duchamp}.

When the merging is completed further criteria are applied which filter out
emission significantly smaller than the beam ($<$7 pixels), or with
unusually narrow linewidths ($<$ 2 channels, i.e., a minimum width of
$0.84\,\kms$). Finally a mask cube is created which uses unique
integers to identify the contiguous emitting regions.

To facilitate the measurement and analysis steps the final mask cube
is used as a template to excise each {\sc duchamp} cloud from the
original HOPS $\nhthree$ data-cubes. Small FITS cubes are produced
containing one cloud each in addition to 2- and 3-D masks indicating
which pixels/voxels contain emission.


\subsection{Emission finder inputs}
The critical inputs for the second pass of {\scriptsize
  DUCHAMP} are presented in Table~\ref{tab:duchamp_inputs}. These are
correct for {\scriptsize DUCHAMP} version 1.1.10. Before
running the finder we performed the {\it \`a trous} reconstruction on
the data as it confers a considerable advantage in detecting compact
and weak emission. The ideal reconstruction parameters are dependant on the
noise properties of the cubes and we experimented with different values
of {\it snrRecon}, {\it scaleMax} and {\it filterCode} (see
Table~\ref{tab:duchamp_inputs}) until the
residuals contained only noise. The search threshold
was based on completeness tests performed using artificial sources
injected into the data (see Section~\ref{sub:src_completeness}). A 0.8$\sigma$ threshold was used, equivalent to
$\sim$\,0.16\,K and sensitive enough to sample the completeness curve
below the one percent level. The last critical parameter is minimum
number of pixels required in a detection. With a beam-size of $\Theta_{\rm\,
  FWHM}$\,=\,2~arcmin and pixel dimensions of
$\Delta_{pix}=0.5$~arcmin, there are 12.6 pixels within the solid
angle subtended by the beam FWHM. For weak and unresolved emission we
may only detect the peak of the beam function, hence we set a
threshold of seven pixels per beam. We justify our input criteria
fully in Section~\ref{sub:src_completeness}.


\label{lastpage}

\end{document}